\documentclass[aps,prl,twocolumn, english, floatfix, 
superscriptaddress, 10pt]{revtex4-2}
\usepackage[T1]{fontenc}
\usepackage[latin1]{inputenc}
\usepackage{lmodern}
\expandafter\let\csname equation*\endcsname\relax
\expandafter\let\csname endequation*\endcsname\relax
\usepackage{amsmath}
\usepackage{txfonts}
\usepackage{esint}
\usepackage{amssymb}
\usepackage{graphicx}
\usepackage{xcolor}
\usepackage{natbib}
\usepackage{bm}
\usepackage{bbold}
\usepackage[cal=boondoxo]{mathalfa}
\usepackage{comment}
\usepackage{ulem} 

\graphicspath{{./img/}}

\def\bra#1{\mathinner{\langle{#1}|}}
\def\ket#1{\mathinner{|{#1}\rangle}}
\def\braket#1{\mathinner{\langle{#1}\rangle}}

\def\pM{\mathrel{\raise 2pt \hbox{\tiny(}\!\raise 1pt \hbox{+}\settowidth {\dimen03} {+}\hskip-\dimen03 \raise -2.4pt \hbox {$-$} \!\raise 2pt \hbox{\tiny)}}}

\renewcommand*{\vec}[1]{{\boldsymbol{\mathrm{#1}}}} 
\newcommand{\1}{\mathbb{1}}

\usepackage[unicode=true,pdfusetitle,
bookmarks=true,bookmarksnumbered=true,bookmarksopen=true,bookmarksopenlevel=2,
breaklinks=false,pdfborder={0 0 0},backref=false,colorlinks=true]{hyperref}
\hypersetup{colorlinks=true, linkcolor=blue, citecolor=blue, urlcolor=blue}

\usepackage{bookmark}
\newcommand*{\heading}[1]{\belowpdfbookmark{#1}{#1}{\bfseries\textit{#1.---}}\ignorespaces}
\let\oldsection\section
\let \section \heading

\begin{document}

\title{
{Charged moir\'e phonons in twisted bilayer graphene}\\}

\newcommand{\Columbia}{Department of Physics, Columbia University, New York, NY 10027, USA}

\author{Alejandro Ramos-Alonso}
\email[]{ar4612@columbia.edu}
\affiliation{\Columbia}

\author{H{\'e}ctor Ochoa}
\email[]{ho2273@columbia.edu}
\affiliation{\Columbia}

\begin{abstract}

Moir\'e phonons describe collective vibrations of a moir\'e superlattice produced by long-wavelength relative displacements of the constituent layers. Despite coming from the backfolding of the acoustic phonons of the individual layers, many of these modes become infrared active when the system is doped. We illustrate this effect by a direct calculation of the optical absorption of twisted bilayer graphene (tBG) around different twist angles, including the magic angle. Several modes --including the acoustic-like phason-- acquire a dipole moment via interband matrix elements of the electron-phonon coupling (EPC) when the flat band is filled or emptied, giving rise to new resonances in the optical conductivity within the single-electron gap that are strongly affected by relaxation. The phason in particular gains a charge that equals the amount of electrons per moir\'e cell added/removed to/from neutrality. Geometrically, this can be understood as the topological quantization of a sliding Chern number.
The charged phason yields a Drude-like conductivity with an effective mass that increases with lattice relaxation. 
Our findings are testable via THz spectroscopy, and provide an experimental knob to characterize EPC strength and disorder in moir\'e materials at small twist angles.

\end{abstract}

\maketitle

\section{Introduction} 
Electrons in moir\'e patterns form flat bands \cite{MacDonald2011, Barticevic2010}, favoring the onset of correlations among them.  
Despite the extensive experimental evidence of correlated behavior, such as superconductivity \cite{Jarillo2018, Efetov2019, Wang2019, Jarillo2021, Jarillo2022,Uri:2023aa,FaiMak2024, Dean2025}, a complete theoretical understanding of these phases is still lacking, in particular, as it pertains to the interplay of moir\'e patterns and their fluctuations (including twist angle disorder) with charge carriers.
Among all the possible assemblies, twisted bilayer graphene (tBG) stands as a homobilayer of reference. However, flat bands in commensurate stacks, like in Bernal bilayers or in rhombohedral trilayers and pentalayers, also lead to superconductivity in proximity to (or within) flavor-polarized states \cite{Zhou:2021aa,doi:10.1126/science.abm8386,Holleis:2025aa,Choi:2025aa,Han:2025aa}. Besides superconductivity, tBG displays strange metal behavior in its normal state \cite{Young2019, Jarillo2020, Efetov2022}, a phenomenon not reported in commensurate stacks.

A distinct feature of twisted Van der Waals layers is that they are plagued by soft phonons introduced by incommensurability. At small twist angles, moir\'e systems relax towards a soliton network of domain walls that separate regions of different stacking \cite{nam2017lattice,carr2018relaxation,LeRoy2018, Basov2018,Kim2023, Edmonds2024,Shaffique2024,KangVafek2025,DeBeule2025}. 
The associated long-wavelength collective modes of the stacking configuration (dubbed \textit{moir\'e phonons} \cite{Koshino2019}) include two acoustic branches (\textit{phasons}) corresponding to the invariance of the free energy with respect to a relative shift of the layers \cite{Ochoa2019}.
The role of moir\'e phonons in the physics of twisted bilayers and their experimental signatures are object of current research \cite{KoshinoNam2020,phasons_TMD,xiao2021chiral,lu2022low,Fernandes2022,Scheurer2022,Eslam2022,Liu2022,VafekKang2023,Girotto2023,Guinea2023,Maity2023,Xiao2024,Ochoa2025}, part of it fueled by its spectral characterization \cite{Birkbeck2025} using the recently developed quantum twisting microscopy \cite{Inbar2023}, as well as its direct observation in real space using high-resolution electron ptychography \cite{Zhang2025}. 
In particular, electron scattering off phasons can lead to strange metal behavior in tBG \cite{Fernandes2023}, a proposal supported by numerical and experimental evidence from  minimally twisted samples \cite{Pezzini2025}.

\begin{figure}[t!]
\begin{center}
\includegraphics[width=\columnwidth]{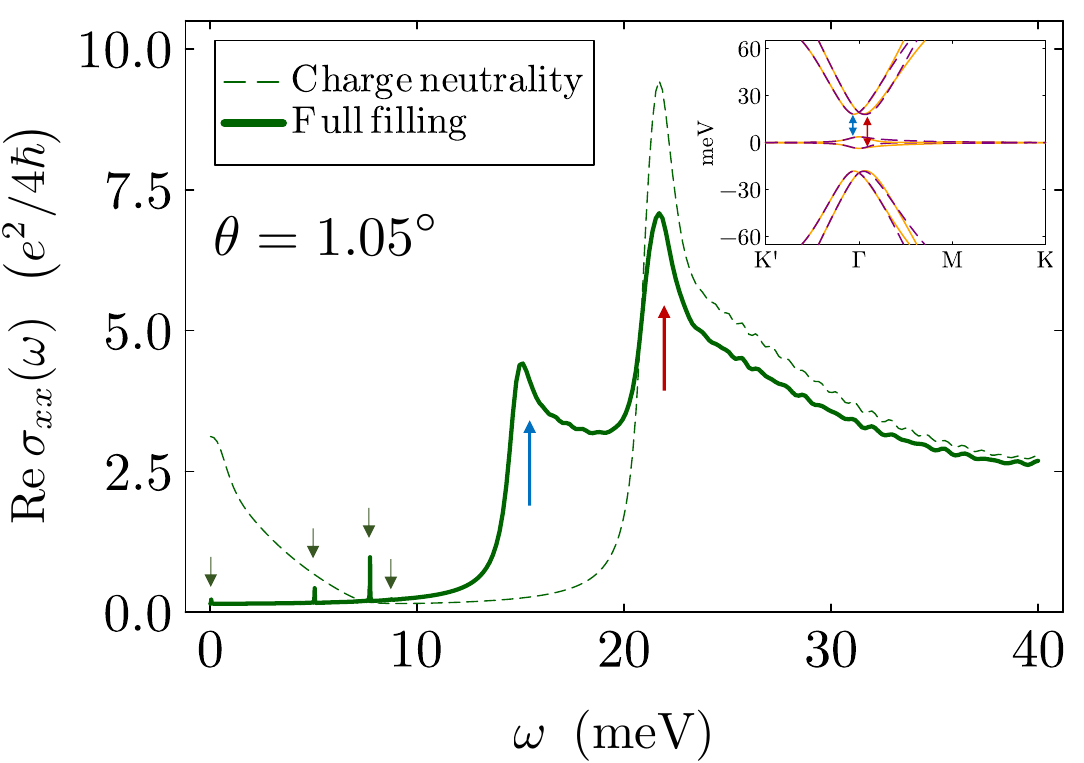}
\caption{Optical absorption of tBG at the magic angle in the absence of interlayer adhesion at two different fillings of the flat bands. Small green arrows highlight the moir\'e phonon peaks. 
The inset shows the electronic band structure, where color distinguishes valleys. The interband transitions giving rise to peaks at finite frequency in the electron-hole continuum are indicated by the blue and red arrows.
}
\label{fig:fig1}
\vspace{-0.3cm}
\end{center}
\end{figure}

\begin{figure*}[t!]
\begin{center}
\includegraphics[width=2.1\columnwidth]{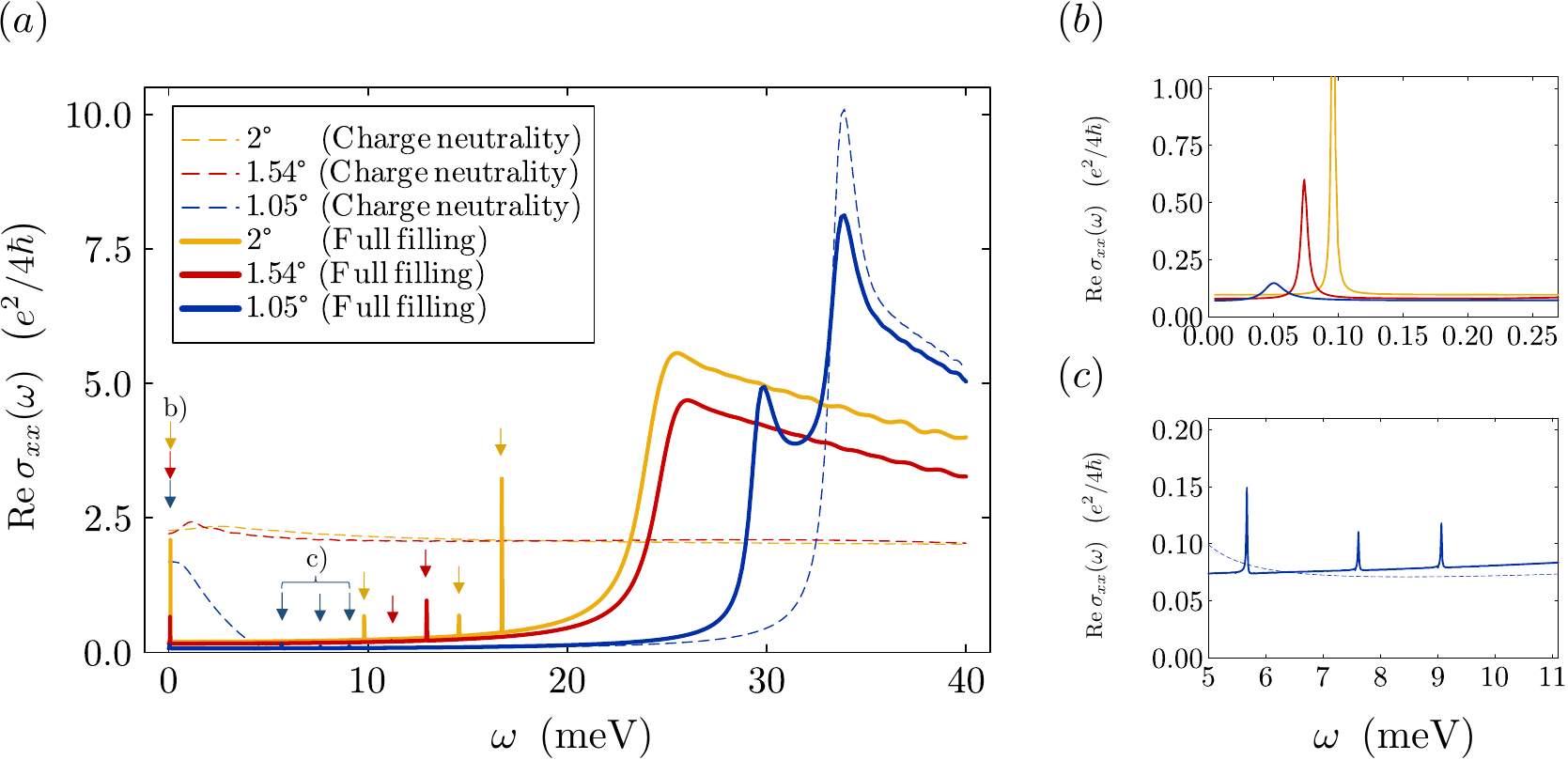}
\caption{ (a) Optical absorption of tBG for the neutral (dashed lines) and for the doped system with four electrons per moir\'e cell (solid lines) at three different twist angles in the presence of interlayer adhesion forces. 
(b) Phason peaks in the optical absorption. For visualization, they are pinned to $0.01\omega_{m}$, where the characteristic phonon frequency imposed by folding scales as $\omega_{m}\propto\theta$ \cite{SI}.
(c) Optical absorption at the magic angle in the range of frequencies of the lowest-energy infrared-active gapped modes.
}
\label{fig:fig2}
\vspace{-0.3cm}
\end{center}
\end{figure*}

In this Letter we show that several moir\'e phonons, including the phason branches, become infrared active when the system is doped, despite the fact that these modes originate from the zone-folding of the monolayer's acoustic phonons, which do not couple directly to light. This effect arises from the inhomogeneous distribution of charge in the alternating stacking domains of moir\'e patterns. The coupled oscillation of stacking and charge endows moir\'e phonons of $E_1$ symmetry with a dipole moment by borrowing optical spectral weight from the electron-hole continuum. This is the phenomenon of \textit{charged phonons} introduced in the context of organic semiconductors \cite{Rice1976,Choi1992} and more recently observed in Bernal bilayer graphene \cite{Kuzmenko2010,Kuzmenko2012,Geim2009}. It is generically present in any charged moir\'e system, but magic-angle tBG is optimal for its detection: As illustrated by the calculation in Figs.~\ref{fig:fig1} and ~\ref{fig:fig2}, several moir\'e phonon resonances lie within the single-electron gap, which is enlarged by the relaxation of the superlattice, when the flat band is fully occupied (or emptied).

The infrared activity of phasons can be understood as an example of a \textit{sliding conductivity} no different from a charge density wave \cite{Anderson1974}. We show explicitly how the phason resonance arises from the combined effect of the adiabatic pumping of charge when one layer slides over the other \cite{Koshino2020,DiXiao2020,Lin2020} together with the reciprocal backaction of charge on the interlayer shear mechanics \cite{Ochoa2023}. This result adds to the list of differences between acoustic phonons of crystalline lattices and acoustic phasons of incommensurate moir\'e superlattices.

\section{Model} 
Moir\'e phonons describe fluctuations in the stacking order produced by long-wavelength shear displacements of the layers. In our model the local stacking configuration around position $\mathbf{r}$ in a homobilayer (in Eulerian coordinates) is described by a vector field $\boldsymbol{\phi}(\mathbf{r})$ representing the lateral relative displacement of a given element on each layer due to a lattice mismatch produced by a twist, heterostrain, or both, taking the case of maximum lattice overlap (AA stacking in bilayer graphene) as the reference ($\boldsymbol{\phi}=0$). By construction, $\boldsymbol{\phi}$ and $\boldsymbol{\phi}+\mathbf{R}$ represent the same stacking configuration, where $\mathbf{R}$ is
a Bravais vector in the reference state. Provided that the layers are stiff and adhesion forces are weak, for small twist angles the stacking field $\boldsymbol{\phi}(\mathbf{r})$ can be assumed to be smooth on the atomic scale; thus we use a continuum model to describe moir\'e phonons and their coupling to electronic degrees of freedom. 

In the absence of free electrons, the energy cost of spatial variations of $\boldsymbol{\phi}(\mathbf{r})$ comes from the elasticity of each layer and from their mutual adhesion interaction. 
Following the methods in Ref. \cite{Ochoa2025}, the resulting functional is minimized to find the equilibrium configuration $\boldsymbol{\phi}_0(\mathbf{r})$, while the dispersion of moir\'e phonons follows from the identification of normal modes --labeled by momentum $\mathbf{k}$ within the moir\'e Brillouin zone (mBZ) and branch index $\mu$-- in the harmonic expansion of this functional around the equilibrium solution, $\delta\boldsymbol{\phi}(\mathbf{r})=\boldsymbol{\phi}(\mathbf{r})-\boldsymbol{\phi}_0(\mathbf{r})$, where\begin{align}
\delta\boldsymbol{\phi}(\mathbf{r})=\frac{1}{\sqrt{A}}\sum_{\mu}\sum_{\mathbf{k}\in\textrm{mBZ}}\sum_{\left\{\mathbf{G}\right\}}\delta\boldsymbol{\phi}_{\mathbf{G}}^{(\mu)}(\mathbf{k})\,\phi_{\mu,\mathbf{k}}\, e^{i\left(\mathbf{k}+\mathbf{G}\right)\cdot\mathbf{r}},
\end{align}
$\mathbf{G}$ being reciprocal vectors of the moir\'e pattern, $\phi_{\mu,\mathbf{k}}$ the new normal-mode coordinates, and $\delta\boldsymbol{\phi}_{\mathbf{G}}^{(\mu)}(\mathbf{k})$ dimensionless coefficients describing the spatial modulation of the displacements associated with mode $(\mu,\mathbf{k})$. The latter are obtained from the diagonalization of the corresponding dynamical matrix and satisfy the normalization condition $\sum_{\{\mathbf{G}\}}|\delta\boldsymbol{\phi}_{\mathbf{G}}^{(\mu)}(\mathbf{k})|^2=1$. 

In the absence of adhesion forces the phonon dispersion comes from folding the original acoustic phonons of graphene back to the mBZ. 
In the presence of adhesion forces the spectrum is reconstructed in new moir\'e phonon branches. 
The effect of the moir\'e potential is larger for smaller angles.
The surviving acoustic branches dispersing down to $\omega=0$ at the zone center can be identified with sliding phason modes \cite{SI}. These, along with several optical modes, form $E_1$ doublets of the $D_6$ point group of tBG at the $\Gamma$ point.
We show the dispersion of moir\'e phonons of tBG for different twist angles \cite{SI}.

Electrons in tBG are described by the continuum model \cite{CastroNeto2007,MacDonald2011,Balents2019}
\begin{align}\label{eqn:H1}
\hat{H}[\boldsymbol{\phi}(\mathbf{r})] =
    \sum_{\xi=\pm 1}\int d\mathbf{r}\: \hat{\psi}^{\dagger}_{\xi}(\mathbf{r})\: \hat{\mathcal{H}}_{\xi}[\boldsymbol{\phi}(\mathbf{r})] \:\hat{\psi}_{\xi}(\mathbf{r}),
\end{align}
where $\hat{\psi}_{\xi}$ are 4-component field operators defined in sublattice and layer spaces representing electronic excitations around graphene's inequivalent valleys $(\xi=\pm1)$, and $\hat{\mathcal{H}}_{\xi}[\boldsymbol{\phi}(\mathbf{r})]$ is the block-matrix Hamiltonian \begin{align}\label{eqn:H2}
    \hat{\mathcal{H}}_{\xi}[\vec{\phi}(\vec{r})] =
    \begin{pmatrix}
    \hat{\mathcal{H}}_{\xi}^{(t)}[\vec{\phi}(\vec{r})] & \hat{\mathcal{T}}_{\xi}[\vec{\phi}(\vec{r})] \\ \hat{\mathcal{T}}_{\xi}^{\dagger}[\vec{\phi}(\vec{r})] & \hat{\mathcal{H}}_{\xi}^{(b)}[\vec{\phi}(\vec{r})]
    \end{pmatrix}.
\end{align}
The diagonal blocks $\hat{\mathcal{H}}_{\xi}^{(t,b)}$ correspond to the Dirac Hamiltonian of the top and bottom layers in which the components of the heterostrain tensor --symmetrized derivatives of $\boldsymbol{\phi}(\mathbf{r})$-- couple to electrons through a scalar (deformation potential) and pseudo-gauge fields \cite{SI}.
The off-diagonal blocks encode interlayer tunneling processes that are sensitive to the local stacking configuration via phase shifts of the form
\begin{align}\label{eqn:T}
    \hat{\mathcal{T}}_{\xi}[\vec{\phi}(\vec{r})] = \sum_{n=0,1,2} e^{i\xi \vec{Q}_n\cdot\vec{\phi}(\vec{r})}\hat{T}_{\xi}^{(n)}.
\end{align}
This equation describes the tunneling of an electron of one layer as the superposition of three plane waves with the same momentum but measured from the three equivalent corners $\vec{Q}_n$ of the Brillouin zone of the other layer. The sublattice structure of these plane waves is contained in the matrices $\hat{T}_{\xi}^{(n)}$, obtained as $2\pi n\xi/3$-rotations of
$\hat{T}_{\xi}^{(0)} := w_{\textrm{AA}}\hat{\1} + w_{\textrm{AB}}\hat{\sigma}_{x}$, where $\hat{\sigma}_x$ is a Pauli matrix acting on those degrees of freedom. In our calculations, we use the values $w_{\textrm{AA}}=0.0797$ eV and $w_{\textrm{AB}}=0.0975$ eV \cite{Koshino2019electrons}.

The electron-(moir\'e) phonon coupling (EPC) deduced from this model follows from the leading-order expansion in the fluctuation around the solution $\boldsymbol{\phi}_0(\mathbf{r})$, while the electron band Hamiltonian is simply $\hat{H}_0\equiv\hat{H}[\boldsymbol{\phi}_0(\mathbf{r})]$. We consider two scenarios: with no adhesion forces, namely, $\boldsymbol{\phi}_0(\mathbf{r})=2\sin\frac{\theta}{2}\,\mathbf{\hat{z}}\times\mathbf{r}$ describing a rigid rotation of the layers, and with the account of these forces, $\boldsymbol{\phi}_0(\mathbf{r})=2\sin\frac{\theta}{2}\,\mathbf{\hat{z}}\times\mathbf{r}+\mathbf{u}_0(\mathbf{r})$, where $\mathbf{u}_0(\mathbf{r})$ is a periodic function on the moir\'e pattern that we determine self-consistently to minimize the mechanical free energy \cite{Ochoa2025}. 
In the first case we just recover the expressions of the usual continuum model. In the second case higher moir\'e harmonics describe a sharper (on the moir\'e scale) stacking texture. We retain only processes coming from the expansion to linear order in $\mathbf{u}_0$, for the relaxation displacements are smaller than the scale set by $|\mathbf{Q}_n|^{-1}$. 
Representative calculations of the electronic bands of each scenario are included in \cite{SI}.

\begin{figure}[t!]
\begin{center}
\includegraphics[width=\columnwidth]{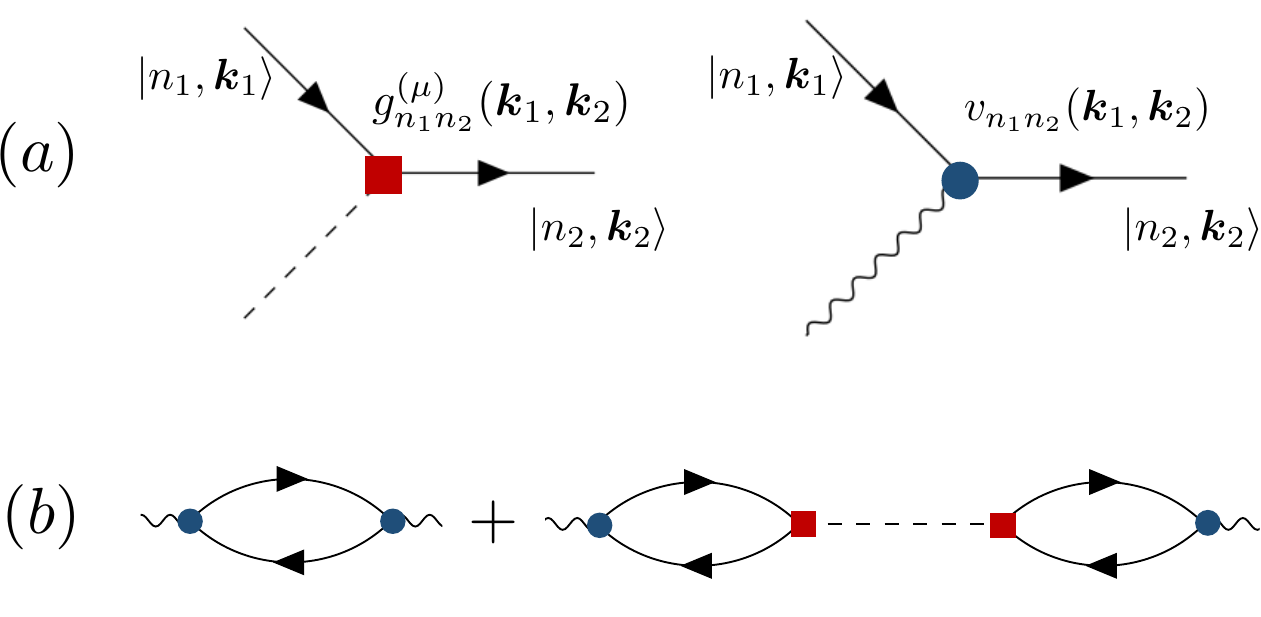}
\caption{Diagrams for the calculation of the optical conductivity. Dashed lines represent phonons, solid lines represent electrons (with band/valley and momentum indices indicated in the figure), and curvy lines correspond to photons.  (a) Electron-phonon (given in Eq.~\ref{eqn:g1}) and electron-photon interaction vertices. (b) Optical conductivity. The first diagram represents the direct current-current response, Eq.~\eqref{eqn:sigma(0)}, while the second diagram involves mixed responses of current and stacking, Eq.~\eqref{eqn:sigma(1)}. 
} 
\label{fig:fig3}
\vspace{-0.3cm}
\end{center}
\end{figure}

\section{Optical conductivity} 
Figure~\ref{fig:fig3} shows the relevant diagrams for the calculation of the optical conductivity. The first vertex in Fig.~\ref{fig:fig3}~(a) (red square) represents the coupling with a moir\'e phonon excitation. The EPC matrix elements in the electron-band and normal-mode basis can be written as
\begin{align}\label{eqn:g1}
    g^{(\mu)}_{n_{1},n_{2}}(\vec{k}_{1},\vec{k}_{2}) = \bra{n_{2},\vec{k}_{2}} \frac{\partial\hat{\mathcal{H}}_{\xi}[\vec{\phi}] }{\partial\phi_{\mu,\vec{k}_{2}-\vec{k}_{1}}}\bigg|_{\vec{\phi}_0(\vec{r})} \ket{n_{1},\vec{k}_{1}},
\end{align}
where the valley index $\xi$ has been incorporated to the electron band indices $n_{1,2}$. The explicit mathematical expression of these matrix elements can be found in \cite{SI}. The second vertex in Fig.~\ref{fig:fig3}~(a) (blue circle) represents the coupling with a photon.

In random phase approximation (RPA) the conductivity can be divided in two terms, $\sigma_{xx}(\omega)=\sigma_{xx}^{(0)}(\omega)+\sigma_{xx}^{(1)}(\omega)$. The first one is related to the direct current-current response depicted in the first diagram of Fig.~\ref{fig:fig3}~(b),
\begin{align}\label{eqn:sigma(0)}
    \sigma_{xx}^{(0)}(\omega) = \underset{\vec{k}\rightarrow 0 }{\lim}\: \frac{i}{\omega}\chi_{j_{x}(\vec{k})j_{x}(-\vec{k})}(\omega). 
\end{align}
It describes the contribution from the (bare) electron-hole continuum. The other term, corresponding to the second diagram in Fig.~\ref{fig:fig3}~(b), represents the RPA moir\'e-phonon correction to the conductivity,
\begin{align}\label{eqn:sigma(1)}
    \sigma_{xx}^{(1)}(\omega) = \underset{\vec{k}\rightarrow 0 }{\lim}\: \frac{i}{\omega} 
    \sum_{\alpha\in E_1}
    \chi_{j_{x}(\vec{k})\phi_{\alpha,y}(-\vec{k})}(\omega) D_{\alpha}(\vec{k},\omega) \chi_{\phi_{\alpha,y}(\vec{k})j_{x}(-\vec{k})}(\omega),
\end{align}
where the bubbles are mixed response functions of currents and stackings, and the dashed line is the phonon propagator,
\begin{align}
    D_{\alpha}(\vec{k},\omega) = \frac{2}{\varrho} \frac{1}{\omega^{2}-\omega^{2}_{\alpha,\vec{k}} + i\gamma\omega}.
\end{align}
Here $\varrho$ is the mass density of one layer, $\omega_{\alpha,\vec{k}}$ is the phonon dispersion, and the factor $2$ appears as we must consider the inertia of the relative motion.
In our calculations, the phenomenological damping $\gamma$ is assumed to scale as $\theta^{-3}$ \cite{Fernandes2022}, and its reference value is set from a fit to transport measurements of minimally twisted tBG \cite{Pezzini2025, SI}.

\begin{figure}[t!]
\begin{center}
\includegraphics[width=\columnwidth]{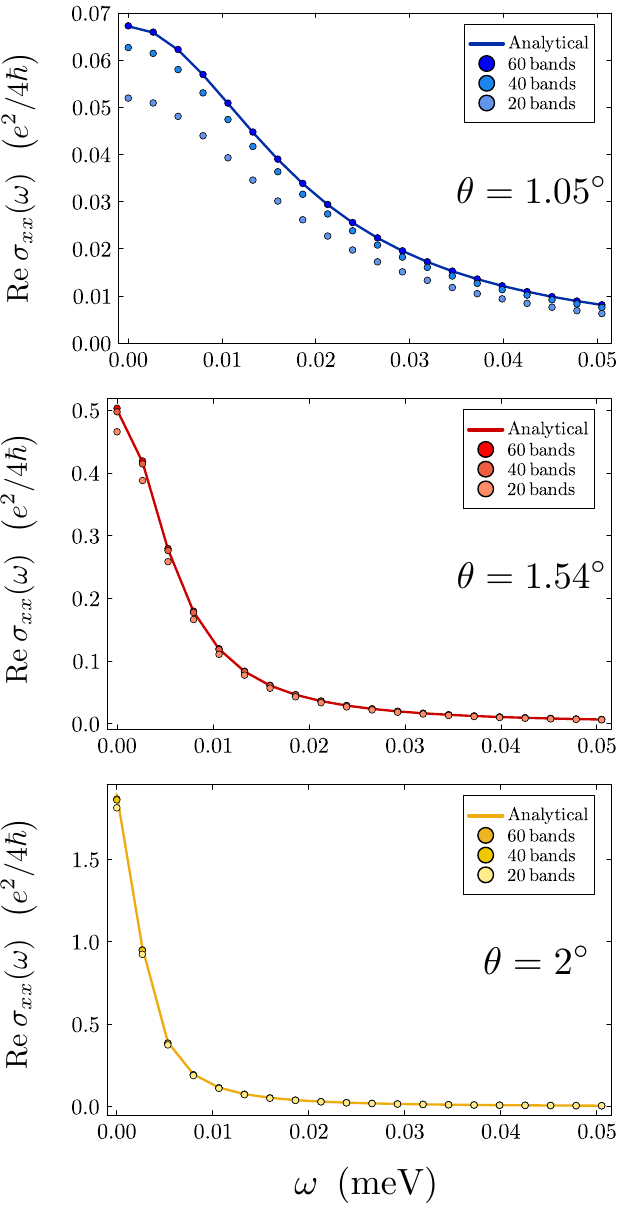}
\caption{Optical absorption of the phason mode with no pinning for the three twist angles of reference. 
The greater the relaxation (i.e. the smaller the twist angle), the more electronic interband transitions are needed in Eq. \eqref{eqn:sigma(1)} in order to reproduce the analytical curve given by Eq. \eqref{eqn:sigmaph}. 
} 
\label{fig:fig4}
\vspace{-0.3cm}
\end{center}
\end{figure}

In the $\mathbf{k}=0$ limit the mixed current-stacking response is nonzero only for modes with the same symmetry as the current operator, hence the summation in Eq.~\eqref{eqn:sigma(1)} is restricted to zone-center phonons forming $E_1$ doublets; this is reflected in the change of notation: $\phi_{\mu,\mathbf{0}}\rightarrow \vec{\phi}_{\alpha}=(\phi_{\alpha,x},\phi_{\alpha,y})$ \cite{SI}. Note that this vector transforms just like $(-j_y,j_x)$ under $D_6$ operations. 
The RPA resummation --justified by the \textit{large} ($N=4$) number of fermion species--  should 
include phonon self-energy corrections due to the coupling with electrons in second-cumulant approximation. We find that their effect on the position of the phonon peaks is  small (e.g., compared to corrections by adhesion forces).  
Moreover, the broadening introduced by the coupling with electrons in RPA approximation vanishes for modes within the optical gap \cite{SI}, hence  our phenomenological $\gamma$ of (assumed) mechanical origin is the only source of damping of moir\'e phonons.
We provide an analytical proof that phasons remain gapless in our theory \cite{SI}, in accordance with Adler's principle \cite{PhysRev.139.B1638,Watanabe2014}.

Figure~\ref{fig:fig1} shows the numerical evaluation of the optical absorption (real part of $\sigma_{xx}$) of magic-angle tBG as a function of frequency in the absence of adhesion forces. 
The dashed line represents the neutral system, when the response is fully accounted for by the electron-hole continuum. 
The continuous line corresponds to a filling of four electrons per moir\'e cell. 
There are two prominent features within the electron-hole continuum coming from interband transitions between the fully occupied flat bands and the next dispersive bands, as indicated in the insets.
Additional resonances appear within the optical gap at the frequencies of the four lowest-energy infrared-active moir\'e phonons, including the phason mode.
In Fig.~\ref{fig:fig2} we show the full calculation with adhesion forces for three representative twist angles. 
At the magic angle, relaxation effects are important to capture the magnitude of the gap and of the moir\'e phonon responses, displayed in panel (c): 
We find that the local phase shifts in Eq.~\eqref{eqn:T} push the dispersive bands to higher energies \cite{SI}, effectively reducing the optical absorption.  

Our expressions exhaust the f-sum rule by virtue of the causality of the response functions involved \cite{SI}.
Numerically, most of the spectral weight that builds the dipole moment of the charged moir\'e phonons is borrowed from the peaked features of the electron-hole continuum. 
In particular, the $E_{1}$ mode giving the most intense optical response in Fig.~\ref{fig:fig1} takes 94\% of its weight from the range of frequencies we show in our plots.

\section{Charged phason} 
The phason peak appearing at the lowest energies has a direct geometrical interpretation. For these modes, the calculation can be performed in new collective coordinates $\vec{\zeta}$ describing position shifts of the relaxed structure, $\vec{\phi}_0(\vec{r})\rightarrow\vec{\phi}_0(\vec{r}-\vec{\zeta})$. At small twist angles, the relation between the normal-mode coordinates in stackings and these new coordinates for phasons is simply \cite{SI}\begin{align}
\label{eq:change_variables}
 \vec{\phi}_{\textrm{ph}} = -\frac{\theta}{\eta}\vec{\hat{z}}\times\vec{\zeta},
\end{align}
where $0<\eta\leq 1$ is a dimensionless numerical factor, 
describing how much the phason shift deviates from a rigid, relative translation of the layers. In our calculations, it takes the form  
\begin{align}
    \eta = \frac{1}{\sqrt{ 1 + \frac{1}{2\theta^{2}}\sum_{\{\vec{G}\}} |\vec{G}|^{2}|\vec{u}_{\vec{G}}|^{2} }},
\end{align}
where we have introduced the decomposition $\vec{u}_{0}(\vec{r}) = \sum_{\{\vec{G}\}} \vec{u}_{\vec{G}}e^{i\vec{G}\vec{r}}$.
In the absence of adhesion forces we have $\eta=1$, and Eq.~\eqref{eq:change_variables} is the geometrical relation between the aforementioned relative motion and a perpendicular shift of the moir\'e pattern amplified by the small twist angle. 
When the lattice relaxes due to adhesion forces the picture is not as straightforward, since now more complex ion rearrangements take place during the sliding motion of stacking domain walls.

The dominant term in the low-frequency (compared to the single-electron gap) optical response of the phason is 
\begin{align}
    \sigma_{xx}^{\text{ph}}(\omega) \approx 
    \frac{i\hbar^{2}e^2\omega\,\eta^2 C^2}{\theta^2} D_{\text{ph}}(\vec{0},\omega), 
\end{align}
with 
\begin{align}
\label{eq:slidingChern}
    C=2\sum_{n}^{\textrm{occ}}\int_{\mathrm{mBZ}}\frac{d^2\vec{k}}{(2\pi)^{2}}\, \Omega_{k_{x},\zeta_{x}}^{(n)}(\vec{k})  = -n,
\end{align}
where the sum is restricted to occupied bands, $n$ is the charge density measured from neutrality,  the factor $2$ comes from spin degeneracy, and the integrand is the sliding Berry curvature related to the pump current induced by layer sliding \cite{Koshino2020,DiXiao2020,Lin2020} and the reciprocal electro-mechanical process \cite{Ochoa2023},
\begin{align}
    \Omega^{(n)}_{k_{i},\zeta_{j}}(\vec{k}) &:= 
    \sum_{n'\neq n} \frac{-2\text{Im}[\bra{n\vec{k}}\frac{\partial \hat{\mathcal{H}}_{\xi}}{\partial k_{i}}\ket{n'\vec{k}}
    \bra{n'\vec{k}}\frac{\partial \hat{\mathcal{H}}_{\xi}}{\partial \zeta_{j}}\ket{n\vec{k}}]}{(\varepsilon_{n',\vec{k}}-\varepsilon_{n,\vec{k}})^{2}}.
\end{align}
As the moir\'e pattern flows, the charge follows adiabatically, thus the coefficient $C$ multiplied by the area of the moir\'e cell must be quantized in units of the number of electrons added/removed to/from the neutral system.  
Numerically, contributions involving bands far apart in energy are non-negligible because of its geometrical origin.
Figure~\ref{fig:fig4} shows the convergence of Eq.~\eqref{eqn:sigma(1)} with the number of bands included in the calculation.
At small angles, accounting explicitly for relaxation in the electronic Hamiltonian through higher moir\'e harmonics (i.e. not only as a correction of the parameters of the model) is pivotal in reproducing the analytical result \cite{SI}.

It follows then that the phason is charged with exactly the amount of charge introduced in the system, and its optical response at small frequencies is given by
\begin{align}\label{eqn:sigmaph}
    \sigma^{\text{ph}}_{xx}(\omega) \approx\frac{n e^2\tau}{m^*}\frac{1}{1-i\tau\left(\omega-\frac{\omega_{\textrm{ph},\vec{0}}^2}{\omega}\right)},
\end{align}
where $\tau=\gamma^{-1}$ is the phason lifetime and $m^*$ is the effective mass of the charged moir\'e pattern,
\begin{align}
\label{eq:m_*}
    m^*=\frac{\theta^2\varrho}{2\eta^2 n}.
\end{align}
In the clean limit, $\omega_{\textrm{ph},\vec{0}}=0$, and the conductivity in Eq.~\eqref{eqn:sigmaph} acquires a Drude-like form. In particular, the charged phason makes tBG display DC conductivity despite being a band insulator at full filling. Like in a sliding charge density wave, in which the ionic distortion travels dragging charge, here an electric field causes a layer-shear mechanical force that triggers the relative motion of the layers, which in turn produces a pumping current. In a real device, this resonance might not be found in DC conductivity but at a finite pinning frequency, $\omega_{\textrm{ph},\vec{0}}\neq0$. For that to happen, it is enough to have any form of disorder \cite{Fernandes2022}, or simply the presence of contacts that prevent the layers from sliding.

\section{Discussion}
We have presented a mechanism from which moir\'e phonons in tBG can get effectively charged when electrons are added to the system.
Stacking fluctuations carry a dipole moment associated with the inhomogeneous distribution of charge, which couples directly to light. Microscopically, this dipole moment arises from the transfer of spectral weight from the electron-hole continuum via interband EPC, giving rise to resonances in the optical conductivity at the frequencies of the lowest-energy $E_{1}$ modes, including the phason. In particular, the geometrical origin of the phason charge is found in terms of a sliding Chern number, which is quantized for an insulating system in the number of electrons per moir\'e cell added/removed to/from neutrality. 

The formation of charged moir\'e phonons is not specific of tBG or the magic angle:
At a different twist angle, the position of the peaks shifts according to a characteristic energy scale $\omega_{m}\propto \theta$ \cite{SI}. 
The miniband gap of the configurations used to illustrate the results is large compared to $\omega_{m}$, 
so they conveniently accommodate several moir\'e phonon resonances.  
Unaffected by its magnitude, the adiabatic arguments describing the charging of the phason branch apply as long as the electron bands remain gapped.
On the contrary, at finite frequencies those branches that do not hybridize with the electron-hole continuum would be more (less) absorptive for smaller (larger) single-electron gaps.

Adhesion forces not only introduce lattice relaxation but also produce a subsequent reconstruction of the moir\'e phonon spectrum. For instance, long-wavelength phasons are no longer rigid translations of the layers, but coherent superpositions of plane waves describing local ionic rearrangements associated with the sliding of domain walls.  
The mode reconstruction induces a further redistribution of the optical spectral weight.   
In Fig.~\ref{fig:fig2}(a) the phason peak in the conductivity  is reduced with respect to Fig.~\ref{fig:fig1} following the analytical formula in
Eq.~\eqref{eqn:sigmaph} as the result of an increase in the effective mass of the charged moir\'e pattern, Eq.~\eqref{eq:m_*}. 
Experimentally, the applied pressure on the sample is the appropriate control parameter to probe these claims, as it tunes the effective adhesion forces between layers.
 
Finally, our results imply that optical measurements in the THz domain can be instrumental in characterizing the amount of disorder in the samples (e.g., via the location of the phason peak), as well as the strength of the EPC in tBG and other moir\'e materials, which can be inferred from the oscillator strength associated with the moir\'e phonon resonances. This offers a complementary strategy to the one of the quantum twisting microscope \cite{Inbar2023}, which can track the phonon dispersion and EPC as a function of twist angle \cite{Birkbeck2025}, but whose resolution in tunneling currents is at present limited to twist angles larger than $6^{\textrm{o}}$.

\section{Acknowledgements} 
We thank Hope M. Bretscher and Felix Sturm for useful discussions, and Dmitri N. Basov for inspiring questions concerning charged phonons.

\let\oldaddcontentsline\addcontentsline \renewcommand{\addcontentsline}[3]{}
\bibliography{references}
\let\addcontentsline\oldaddcontentsline

\clearpage
\newpage

\appendix
\onecolumngrid

\let\section\oldsection

\setcounter{secnumdepth}{6}
\setcounter{equation}{0}
\setcounter{figure}{0}
\setcounter{page}{1}
\counterwithout{equation}{section} 

\renewcommand{\theequation}{S\arabic{equation}}
\renewcommand{\thefigure}{S\arabic{figure}}
\renewcommand{\thetable}{S\arabic{table}}
\renewcommand{\thesection}{S\arabic{section}}
\renewcommand{\thesubsection}{\Alph{subsection}}
\renewcommand{\thesubsubsection}{\Roman{subsubsection}}
\renewcommand{\appendixname}{}        

\begin{center}
    {\Large\bf Supplemental Information: Charged moir\'e phonons in twisted bilayer graphene} 
\end{center}
\vspace{1.5em}

\begin{center}
\begin{minipage}{0.8\textwidth}
    \setlength{\parindent}{1em}   
    We outline how to obtain the phonon eigenvector of a $E_1$ irreducible representation, and we show the equivalence of acoustic moir\'e phonons and phasons, the latter understood as a translation of the coordinates of the stacking field.
    An explicit derivation of the electron-phonon coupling (EPC) vertices employed in the main text is provided, as well as an analytical proof of the stability of the phason in the presence of EPC.
    We also prove that our equations respect the optical f-sum rule.
    Finally, we detail the numerical parameters used to produce the figures.
\end{minipage}
\end{center}
\vspace{1.5em}

\tableofcontents

\section{Moir\'e phonons}
\subsection{$E_{1}$ irreducible representations at $\Gamma$}
\begin{figure*}[t!]
\begin{center}
    \includegraphics[width=0.7\columnwidth]{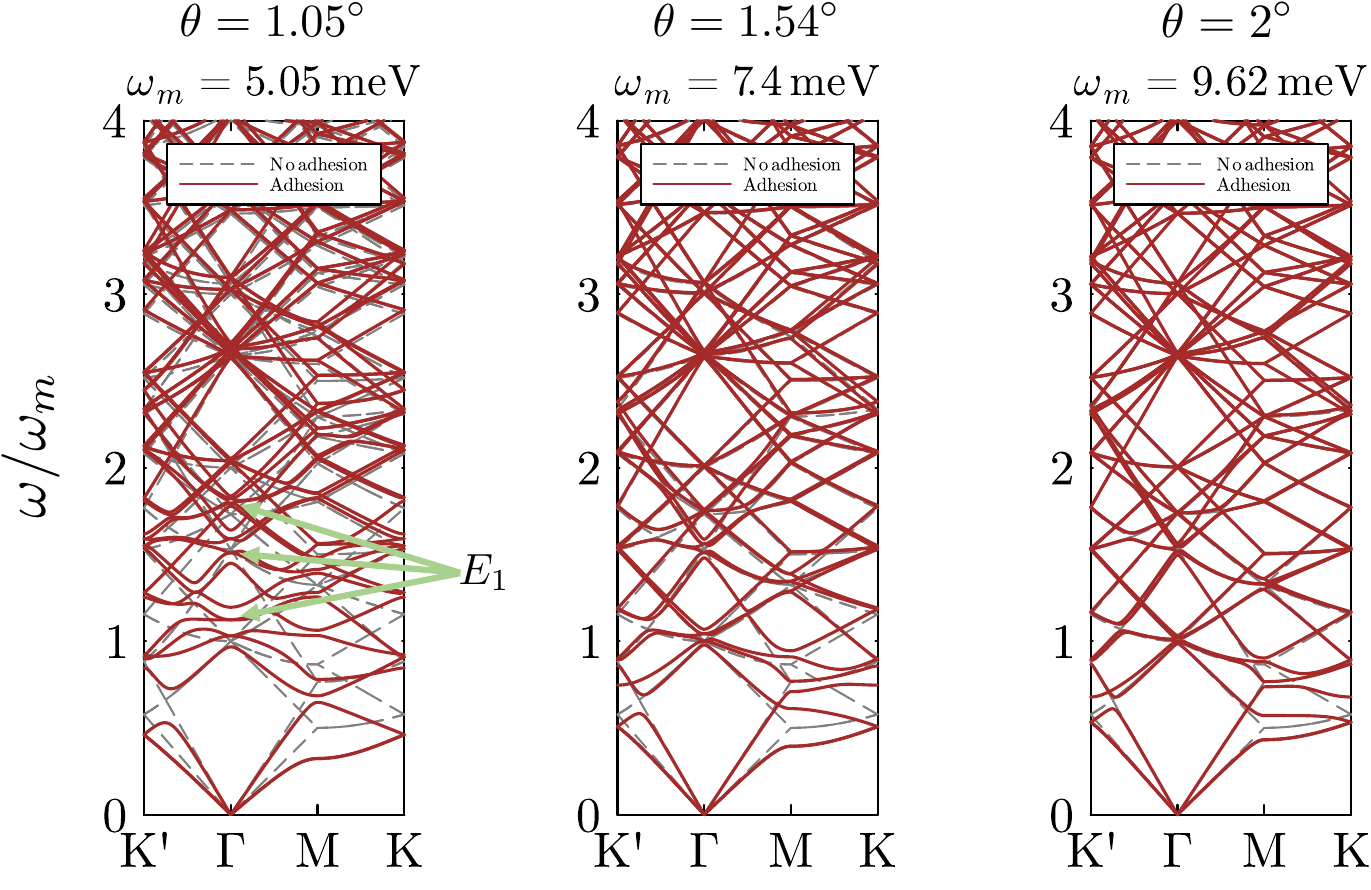}
    \caption{Moir\'e phonon dispersion at different twist angles. The characteristic energy scale is $\omega_{m} = \sqrt{\frac{\mu}{3\varrho}}\frac{4\pi}{L_{m}}\propto\theta$, where $\mu = 9.57$ eV/\AA$^{2}$ and $\varrho = 7.55 \cdot 10^{-27}$ kg/\AA$^{2}$ are the graphene shear modulus and mass density, respectively. $L_{m}$ is the moir\'e periodicity.
    We indicate with green arrows the first three $E_{1}$-branches for $\theta=1.05^{\circ}$. For the other two angles the irreducible representations are ordered in the same way.
    }
    \label{fig:figS1}
    \vspace{-0.3cm}
\end{center}
\end{figure*}

In the absence of adhesion forces, the moir\'e phonon spectrum results from the folding of the monolayer's acoustic phonon dispersion onto the moir\'e Brillouin zone (mBZ). 
As the continuous spatial symmetry is preserved, longitudinal and transverse modes form sixfold degeneracies at $\vec{\Gamma}$. 
In contrast, the moir\'e potential sets the point group of the model to $D_{6}$, whose character table contains a two-dimensional $E_{1}$ irreducible representation, as we indicate in Fig.~\ref{fig:figS1}. In this section we discuss how to form the basis for this representation that we use in our conductivity calculation.

Under a $C_{nz}$ rotation (with $n\in\{2,3 \}$), a stacking fluctuation transforms as $[C_{nz}\delta\vec{\phi}](\vec{r}) = R_{2\pi/n}\cdot\delta\vec{\phi}(R^{-1}_{2\pi/n}\vec{r})$. Here we use $R_{\vartheta}$ to denote rotations of angle $\vartheta$ of the Cartesian coordinates around the $\vec{\hat{z}}$ axis. On the other hand, under a two-fold rotation $C_{2x}$ exchanging the layers, we have $[C_{2x}\delta\vec{\phi}](\vec{r}) = -R_{2x}\cdot\delta\vec{\phi}(R^{-1}_{2x}\vec{r})$, where\begin{align}
    R_{2x}:= \begin{pmatrix}1 & 0 \\ 0 & -1\end{pmatrix}.
\end{align}

A generic fluctuation associated with a $E_1$ mode at $\vec\Gamma$ labeled by index $\alpha$ can be expressed as a linear combination $\delta\vec{\phi}_{\alpha}(\vec{r}) = \sum_{i=x,y} \delta\vec{\phi}_{\alpha,i}(\vec{r}) \phi_{\alpha,i}$, where $\delta\vec{\phi}_{\alpha,i}$ with $i=x,y$ form a representation basis for the $E_1$ doublet, $\{\phi_{\alpha,x}, \phi_{\alpha,y}\}$.   
Using the same plane-wave expansion as in the main text, the Fourier components of the basis elements $\delta\vec{\phi}_{\alpha,i}$ must satisfy:
\begin{align}
    C_{2z} \text{ symmetry: }
    \delta\vec{\phi}_{\alpha,i,\vec{G}}(\vec{\Gamma}) = \delta\vec{\phi}_{\alpha,i,-\vec{G}}(\vec{\Gamma}).
\end{align}
\begin{align}
    C_{3z} \text{ symmetry: }
    &-\frac{1}{2}\delta\vec{\phi}_{\alpha,x,\vec{G}}(\vec{\Gamma}) + \frac{\sqrt{3}}{2}\delta\vec{\phi}_{\alpha,y,\vec{G}}(\vec{\Gamma})
    = R_{2\pi/3}\delta\vec{\phi}_{\alpha,x,R_{2\pi/3}^{-1}\vec{G}}(\vec{\Gamma}),
    \\
    &-\frac{\sqrt{3}}{2}\delta\vec{\phi}_{\alpha,x,\vec{G}}(\vec{\Gamma}) - \frac{1}{2}\delta\vec{\phi}_{\alpha,y,\vec{G}}(\vec{\Gamma})
    = R_{2\pi/3}\delta\vec{\phi}_{\alpha,y,R_{2\pi/3}^{-1}\vec{G}}(\vec{\Gamma}).
\end{align}
\begin{align}\label{eqn:reflection}
    C_{2x} \text{ symmetry: }\delta\vec{\phi}_{\alpha,x,\vec{G}}(\vec{\Gamma}) =R_{2x} \delta\vec{\phi}_{\alpha,x,R_{2x}\vec{G}}(\vec{\Gamma}),
    &&\delta\vec{\phi}_{\alpha,y,\vec{G}}(\vec{\Gamma}) =- R_{2x} \delta\vec{\phi}_{\alpha,y,R_{2x}\vec{G}}(\vec{\Gamma}).
\end{align}

\subsection{Acoustic moir\'e phonons as generators of infinitesimal symmetry transformations}
Consider the equilibrium stacking field $\vec{\phi}_{0}(\vec{r})$. 
We want to see under what conditions a uniform translation of the soliton's coordinates by an arbitrarily-small $\vec{\zeta}$, namely (sum over repeated indices)
\begin{align}\label{eqn:soliton_coords}
    \vec{\phi}_{0}(\vec{r})= \theta\,\vec{\hat{z}}\times\vec{r} + \vec{u}_{0}(\vec{r})  
    \mapsto   
    \vec{\phi}_{0}(\vec{r}-\vec{\zeta}) \approx  
    \vec{\phi}_{0}(\vec{r})  -\theta\vec{\hat{z}}\times\vec{\zeta} - \zeta_{j}\partial_{j}\vec{u}_{0}(\vec{r}), 
\end{align}
can be generated by the two acoustic moir\'e phonons at $\vec{\Gamma}$. Following the notation introduced above, and fixing $\alpha\mapsto\text{ac}$ to denote the lowest-energy doublet,
\begin{align}\label{eqn:realspace}
    \vec{\phi}_{0}(\vec{r}) -\theta\vec{\hat{z}}\times\vec{\zeta} - \zeta_{j}\partial_{j}\vec{u}_{0}(\vec{r})  \overset{?}{=} \vec{\phi}_{0}(\vec{r}) + \delta\vec{\phi}(\vec{r}),
    \text{ such that } 
    \delta\vec{\phi}(\vec{r}) = \sum_{i=x,y}\sum_{ \{ \vec{G} \} }\delta\vec{\phi}_{\text{ac},i,\vec{G}} \phi_{\text{ac},i}e^{i\vec{G}\vec{r}}.
\end{align}
In reciprocal space the problem reads
\begin{align}
    -\theta\vec{\hat{z}}\times\vec{\zeta}\delta_{\vec{G},\vec{0}} + i(\vec{\zeta}\vec{G}) [\vec{u}_{\vec{G}}]^{*} &= 
    \sum_{i=x,y}
    \delta\vec{\phi}_{\text{ac},i,-\vec{G}}
    \phi_{\text{ac}, i},
    \label{eq:system}
\end{align}
where we have used $\vec{u}_{0}(\vec{r}) = \sum_{ \{ \vec{G} \} }\vec{u}_{\vec{G}}e^{i\vec{G}\vec{r}}$.
When adhesion is not considered, $\delta\vec{\phi}_{\text{ac}, i,\vec{G}} = 0$ for all $\vec{G}\neq\vec{0}$, hence $\{\delta\vec{\phi}_{\text{ac},x,\vec{0}},\delta\vec{\phi}_{\text{ac},y,\vec{0}}\} = \left\{\begin{pmatrix}1\\ 0\end{pmatrix} , \begin{pmatrix}0\\ 1\end{pmatrix} \right\}$. 
It follows that
\begin{align}\label{eqn:condition_noadh}
    \begin{pmatrix}\phi_{\text{ac},x}\\ \phi_{\text{ac},y}\end{pmatrix} = \theta\begin{pmatrix}\zeta_{y}\\ -\zeta_{x}\end{pmatrix}
    .
\end{align}

On the other hand, when the interlayer potential is included, the Eq.~\eqref{eq:system} yields the system
\begin{align}\label{eqn:problem}
    \delta\vec{\phi}_{\text{ac},x,\vec{0}}
    \phi_{\text{ac},x} + \delta\vec{\phi}_{\text{ac},y,\vec{0}}
    \phi_{\text{ac},y}   = -\theta\vec{\hat{z}}\times\vec{\zeta}
    ,
\end{align}
\begin{align}\label{eqn:non0}
    \forall\:\vec{G}\neq\vec{0}\quad
    \delta\vec{\phi}_{\text{ac},x,-\vec{G}}
    \phi_{\text{ac},x} + \delta\vec{\phi}_{\text{ac},y,-\vec{G}}
    \phi_{\text{ac},y}
    = 
    i(\vec{\zeta}\vec{G})[\vec{u}_{\vec{G}}]^{*}
    .
\end{align}
In this case,  $\{\delta\vec{\phi}_{\text{ac},x,\vec{0}}, \delta\vec{\phi}_{\text{ac},y,\vec{0}}\} = \left\{\begin{pmatrix}\eta\\ 0\end{pmatrix} , \begin{pmatrix}0\\ \eta\end{pmatrix} \right\}$.
Note that $\eta\neq 0$ is uniquely defined from the diagonalization of the dynamical matrix and the normalization condition $\sum_{ \{ \vec{G} \} }|\delta\vec{\phi}_{\text{ac},i,\vec{G}}|^{2}=1$. Thus,  Eq. \eqref{eqn:problem} holds if and only if
\begin{align}\label{eqn:cond_adh}
    \begin{pmatrix}
        \phi_{\text{ac},x}\\ \phi_{\text{ac},y}
    \end{pmatrix}
    = \frac{\theta}{\eta}
    \begin{pmatrix}
        \zeta_{y}\\ -\zeta_{x}
    \end{pmatrix}.
\end{align}
The additional factor $\eta^{-1}$ compared to Eq. \eqref{eqn:condition_noadh} introduces adhesion-dependence into the phason's effective mass, as discussed in section S3.

In particular, 
\begin{align}
    \zeta_{x}=0 \Longrightarrow \forall\vec{G}\neq \vec{0} \quad
    i\delta\vec{\phi}_{\text{ac},x,-\vec{G}} = -\frac{\eta}{\theta}[\vec{u}_{\vec{G}}]^{*}G_{y},
\end{align}\vspace{-20pt}
\begin{align}\label{eqn:S12}
    \zeta_{y}=0 \Longrightarrow \forall\vec{G}\neq \vec{0} \quad
    i\delta\vec{\phi}_{\text{ac},y,-\vec{G}} = \frac{\eta}{\theta}[\vec{u}_{\vec{G}}]^{*}G_{x},
\end{align}
which we have checked numerically.
Equation~(10) in the main text follows from these relations and the aforementioned normalization condition.

\begin{figure*}[t!]
\begin{center}
    \includegraphics[width=\columnwidth]{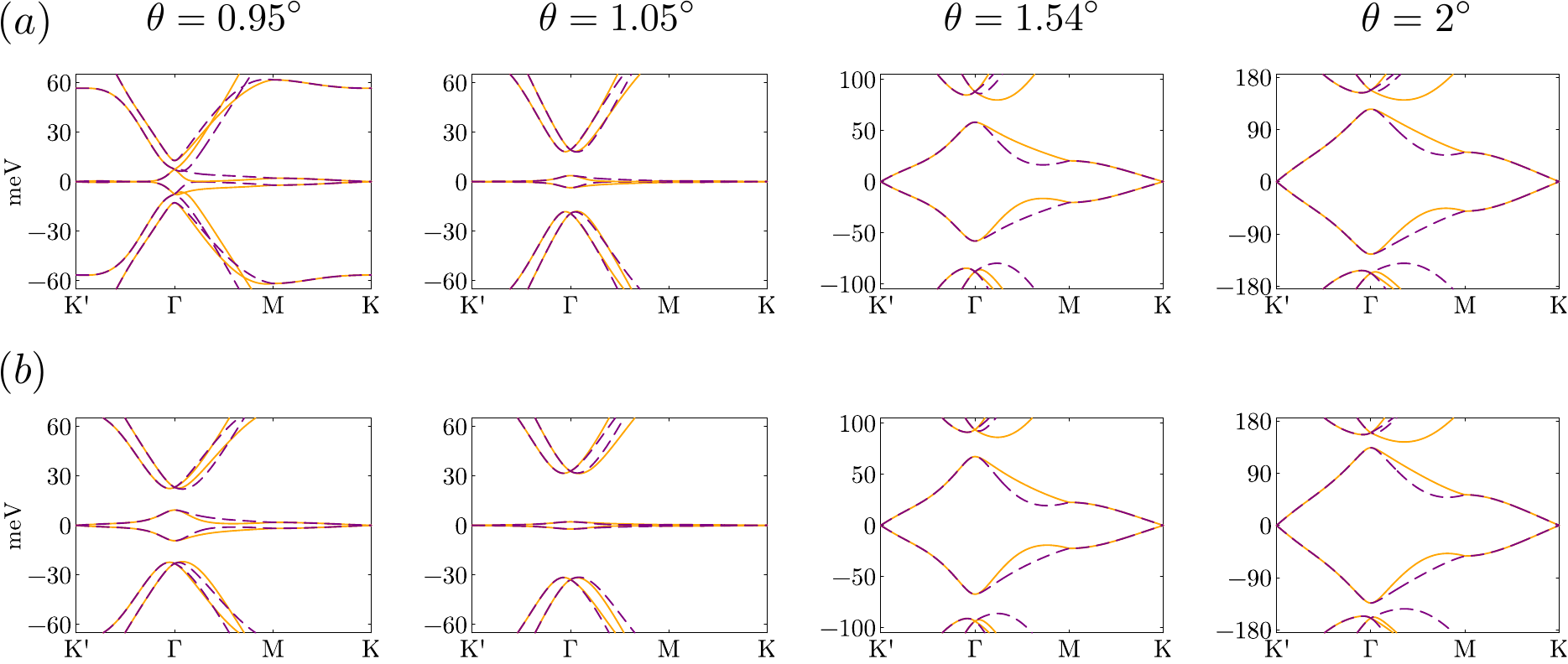}
    \caption{Electronic band structure at different twist angles (a) in the absence of relaxation, (b) with relaxation effects.
    Interlayer adhesion forces are necessary to isolate the flat bands at angles smaller than the magic angle (left column).
    }
    \label{fig:figS2}
    \vspace{-0.3cm}
\end{center}
\end{figure*}

\section{Electron-(moir\'e) phonon coupling (EPC)}
\subsection{EPC matrix elements}
The stacking field can be expressed as the combined deformation of the bilayer by a rigid twist and additional in-plane distorsions $\vec{u}(\vec{r})$, namely 
\begin{align}
    \vec{\phi}(\vec{r}) = \theta\,\vec{\hat{z}}\times\vec{r} + \vec{u}(\vec{r}).
\end{align}
We let $\vec{u}_{0}(\vec{r})$ denote the relaxation field that minimizes the mechanical free energy \cite{Ochoa2025}, and $\delta\vec{\phi}(\vec{r})$ the local fluctuations around the equilibrium stacking configuration $\vec{\phi}_{0}(\vec{r})$. 
In order to study EPC in perturbation theory, we expand the electronic Hamiltonian to second order in the fluctuation:
\begin{align}
    \hat{H}[\vec{\phi}(\vec{r})] = \hat{H}[\vec{\phi}_{0}(\vec{r}) + \delta\vec{\phi}(\vec{r})] \approx \hat{H}[\vec{\phi}_{0}(\vec{r})] 
    + \int d\vec{r}  \frac{\delta\hat{H}[\vec{\phi}_{0}(\vec{r})]}{\delta\vec{\phi}(\vec{r})}\delta\vec{\phi}(\vec{r})
    + \frac{1}{2!}  \int d\vec{r}\int d\vec{r}' \frac{\delta^{2}\hat{H}[\vec{\phi}_{0}(\vec{r})]}{\delta\vec{\phi}(\vec{r}') \: \delta\vec{\phi}(\vec{r})}\delta\vec{\phi}(\vec{r})\delta\vec{\phi}(\vec{r}').
\end{align}
The equilibrium Hamiltonian is originally written in a basis of plane waves, where the electronic wavefunctions of layer $\ell=\pm 1$ (positive sign corresponds to the top layer), sublattice $\alpha$, and valley $\xi=\pm1$ take the form $\psi^{\ell,\alpha}_{\xi,\vec{k}}(\vec{r}) = \frac{1}{\sqrt{A}}\sum_{ \{ \vec{G} \} }\varphi_{\xi,\vec{k},\vec{G}}^{\ell,\alpha}e^{i(\vec{k}+\vec{G} - \xi R_{\ell \theta/2}\vec{Q})\vec{r}}$, where $\vec{G}$ are reciprocal vectors of the moir\'e superlattice and we restrict $\vec{k}\in$ mBZ. 
In ket notation, the basis is $\{\ket{\xi,\vec{k},\vec{G},\ell,\alpha}\}$, such that 
\begin{align}
    \ket{\psi^{\ell,\alpha}_{\xi,\vec{k}}} = \sum_{ \{ \vec{G} \} }\varphi_{\xi,\vec{k},\vec{G}}^{\ell,\alpha}
    \ket{\xi, \vec{k},\vec{G},\ell,\alpha},
    &&
    \braket{\vec{r} | \xi, \vec{k},\vec{G},\ell,\alpha} = \frac{1}{\sqrt{A}} e^{i(\vec{k}+\vec{G} - \xi R_{\ell \theta/2}\vec{Q})\vec{r}}.
\end{align}
The relation with the eigenbasis $\ket{n,\xi,\vec{k}}$ of $\hat{\mathcal{H}}[\vec{\phi}_{0}(\vec{r})]$ is determined by a set of complex coefficients $\{\mathcal{U}_{n}^{\vec{G},\ell,\alpha}(\xi,\vec{k}) \}$:
\begin{align}
    \ket{n,\xi,\vec{k}} = 
    \sum_{ \{ \vec{G} \} }\sum_{\alpha}\sum_{\ell=\pm 1}\mathcal{U}_{n}^{\vec{G},\ell,\alpha}(\xi,\vec{k}) \ket{\xi,\vec{k},\vec{G},\ell,\alpha}
    .
\end{align}
We show in Fig.~\ref{fig:figS2} the band structures that result from diagonalizing the Hamiltonian at different twist angles in the presence or absence of relaxation terms .
To shorten the equations, we separate the contributions to the one-phonon scattering processes of electrons in each valley coming from the pseudogauge potential $\vec{\mathcal{A}}$, the deformation potential $\mathcal{V}$, and the interlayer hopping $\hat{\mathcal{T}}$, respectively:
\begin{align}
    g_{n_{1},n_{2}}^{(\mu)}(\vec{k}_{1},\vec{k}_{2}) = 
    \sum_{\xi=\pm 1} g_{n_{1},n_{2}}^{(\mu,\xi)}(\vec{k}_{1},\vec{k}_{2})[\vec{\mathcal{A}}] +
    g_{n_{1},n_{2}}^{(\mu,\xi)}(\vec{k}_{1},\vec{k}_{2})[\mathcal{V}] + 
    g_{n_{1},n_{2}}^{(\mu,\xi)}(\vec{k}_{1},\vec{k}_{2})[\hat{\mathcal{T}}].
\end{align}
Regarding moir\'e phonons, we keep the convention used in the main text:
\begin{align}\label{eqn:fluc}
    \delta\vec{\phi}(\vec{r}) 
    = \frac{1}{\sqrt{A}}\sum_{\vec{k}\in \text{mBZ}}\sum_{ \{ \vec{G} \} }\sum_{\mu}\delta\vec{\phi}^{(\mu)}_{\vec{G}}(\vec{k}) \phi_{\mu,\vec{k}} e^{i(\vec{k}+\vec{G})\vec{r}}.
\end{align} 
Note that we have imposed periodic boundary conditions $\phi_{\mu,\vec{k}+\vec{G}} = \phi_{\mu,\vec{k}}$ and $\delta\phi^{(\mu)}_{\vec{G}}(\vec{k}+\vec{G}') = \delta\phi^{(\mu)}_{\vec{G}+\vec{G}'}(\vec{k})$.
In what follows, $\Bar{\alpha}:=-\alpha$ and $\Bar{\ell}:=-\ell$.

\subsubsection{Intralayer EPC}
In each graphene layer, electrons move according to a hopping parameter  assumed to decay exponentially with distance:  $\sim e^{-\beta |\vec{r}-\vec{r}\:'|/\sqrt{3}d}$ , where $d=1.425$ $\AA$ is the interatomic distance in graphene and $\beta = 2.5$ is the Gr\"uneisen parameter. The Fermi velocity is $v_{F}=10^{6}$ m/s. To linear order in $\vec{u}$, a pseudogauge field $\vec{\mathcal{A}}$ emerges, such that
\begin{align}
    \vec{\mathcal{A}}[\vec{u}(\vec{r})] = \frac{\beta}{2\sqrt{3}d}
    \begin{pmatrix} -\partial_{x}u_{x}+\partial_{y}u_{y} \\ \partial_{x}u_{y}+\partial_{y}u_{x}\end{pmatrix}.
\end{align}
Additionally, local distorsions introduce a deformation potential $\mathcal{V}=20$ eV term in the electronic Hamiltonian, which reads
\begin{align}
    \hat{\mathcal{H}}_{\xi}^{(\ell)}[\vec{\phi}(\vec{r})] 
    = 
    \hbar v_{F}\vec{\sigma}_{\xi}\cdot \left(-i\nabla +  \ell\xi \boldsymbol{\mathcal{A}}[\boldsymbol{\phi}(\boldsymbol{r})]
    \right)
    +\ell \frac{\mathcal{V}}{2} \left(\partial_i\phi_i\right)\1,
\end{align}
where $\vec{\sigma}_{\xi} = (\xi\sigma_{x},-\sigma_{y})$ contains Pauli matrices acting on the sublattice degree of freedom. We write the resulting intralayer EPC strength in terms of the matrix elements
\begin{align}
    \mathcal{S}_{ij}(\mu,\vec{k},\vec{G}) = (\vec{k}-\vec{G})_{i} \delta\phi^{(\mu)}_{j,-\vec{G}}(\vec{k}), \quad i,j\in\{x,y\}.
\end{align}
In each valley,
\begin{align}
\begin{split}
    g_{n_{1},n_{2}}^{(\mu,\xi)}&(\vec{k}_{1},\vec{k}_{2})[\vec{\mathcal{A}}] = 
    \\ &=
    \sum_{\alpha}\sum_{\ell=\pm 1}\sum_{\vec{G},\vec{G}',\vec{G}''} \sum_{n=0,1,2} 
    i\ell\xi\frac{\beta \hbar v_{F}}{2\sqrt{3}}
    [\mathcal{U}_{n_{2}}^{\vec{G}',\ell,\alpha}(\xi,\vec{k}_{2})]^{*}\: 
    \left[\vec{\sigma}_{\xi}\cdot
    \begin{pmatrix}-\mathcal{S}_{xx}+\mathcal{S}_{yy}\\ \mathcal{S}_{xy}+\mathcal{S}_{yx}\end{pmatrix}\bigg|_{\mu,\vec{k}_{2}-\vec{k}_{1},\vec{G}''}
    \right]_{\alpha,\Bar{\alpha}}
    \mathcal{U}_{n_{1}}^{\vec{G},\ell,\Bar{\alpha}}(\xi,\vec{k}_{1})\:
     \delta_{\vec{G}-\vec{G}'',\:\vec{G}'},
\end{split}
\end{align} 
\begin{align}
    g_{n_{1},n_{2}}^{(\mu,\xi)}(\vec{k}_{1},\vec{k}_{2})[\mathcal{V}] = 
    \sum_{\alpha}\sum_{\ell=\pm 1}\sum_{\vec{G},\vec{G}',\vec{G}''} \sum_{n=0,1,2} 
    i\ell\frac{\mathcal{V}}{2}
    [\mathcal{U}_{n_{2}}^{\vec{G}',\ell,\alpha}(\xi,\vec{k}_{2})]^{*}\: 
    \mathcal{U}_{n_{1}}^{\vec{G},\ell,\alpha}(\xi,\vec{k}_{1})\:
    \text{Tr}[\mathcal{S}(\mu,\vec{k}_{2}-\vec{k}_{1},\vec{G}'')]
     \delta_{\vec{G}-\vec{G}'',\:\vec{G}'}.
\end{align}
Note that the conservation of crystal momentum is taken care of in the evaluation of $\mathcal{S}$.

\subsubsection{Interlayer EPC}
\begin{figure*}[t!]
\begin{center}
    \includegraphics[width=\columnwidth]{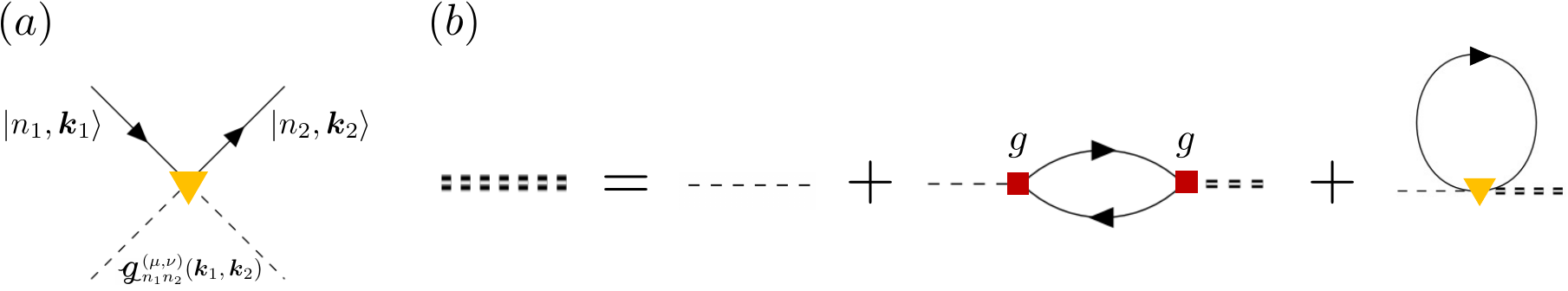}
    \caption{ Diagrammatics of phonon self-energy. (a) Two-phonon vertex, as in Eq. \eqref{eqn:twophonon}. (b) Dyson equation for the phonon renormalization in the one-loop approximation. The dashed double line represents the renormalized phonon propagator.
    The first (second) self-energy diagram is $\Pi_{\text{para}}$ ($\Pi_{\text{diamag}}$).
    }
    \label{fig:figS3}
    \vspace{-0.3cm}
\end{center}
\end{figure*}

Given graphene reciprocal lattice vectors $\vec{g}_{n}$ (set $\vec{g}_{0}=\vec{0}$, $\vec{g}_{1}= \frac{2\pi}{3d}(\sqrt{3}, 1)$, and $\vec{g}_{2}=\frac{2\pi}{3d}(-\sqrt{3}, 1)$) and a Dirac point $\vec{Q}= (\vec{g}_{1}-\vec{g}_{2})/3$, 
define $\vec{q}_{0} = R_{-\theta/2}\vec{Q} - R_{\theta/2}\vec{Q}$ and $\vec{G}_{n} = R_{-\theta/2}\vec{g}_{n} - R_{\theta/2}\vec{g}_{n}$.
It follows that $(\vec{g}_{n}+\vec{Q})\vec{\phi}(\vec{r}) = (\vec{G}_{n}+\vec{q}_{0})\vec{r} + (\vec{g}_{n}+\vec{Q})\vec{u}(\vec{r})$.
The expansion to second order in $\vec{u}_{0}(\vec{r})$ and $\delta\vec{\phi}(\vec{r})$ of the interlayer tunneling Hamiltonian defined in Eq. (4) of the main text thus reads
\begin{align}
    \hat{\mathcal{T}}_{\xi}[\vec{\phi}(\vec{r},t)] \approx \sum_{n=0,1,2} 
     e^{i\xi(\vec{G}_{n}+\vec{q}_{0})\vec{r}}
    \left\{1 + i\xi(\vec{g}_{n} + \vec{Q})(\vec{u}_{0}(\vec{r}) + \delta\vec{\phi}(\vec{r},t)) - \frac{1}{2}[(\vec{g}_{n} + \vec{Q})(\vec{u}_{0}(\vec{r}) + \delta\vec{\phi}(\vec{r},t))]^{2}\right\} 
   \hat{T}_{\xi}^{(n)}.
\end{align}
Two types of scattering events are contained in this expansion. On the one hand, the interaction strength of the one-phonon process is 
\begin{align}\label{eqn:intercoupling}
\begin{split}
    &g_{n_{1},n_{2}}^{(\mu,\xi)}(\vec{k}_{1},\vec{k}_{2})[\hat{\mathcal{T}}] = 
    \\ &=
    \sum_{\alpha,\alpha'}\sum_{\ell=\pm 1}\sum_{\vec{G},\vec{G}',\vec{G}''} \sum_{n=0,1,2}  
    [\mathcal{U}_{n_{2}}^{\vec{G}',\ell,\alpha}(\xi,\vec{k}_{2})]^{*}\: 
    [\hat{T}_{\xi}^{(n)}]_{\alpha,\alpha'}
    \mathcal{U}_{n_{1}}^{\vec{G},\Bar{\ell},\alpha'}(\xi,\vec{k}_{1}) 
    \:
    (\vec{g}_{n}+\vec{Q})\delta\vec{\phi}^{(\mu)}_{-\vec{G}''}(\vec{k}_{2}-\vec{k}_{1})
    \:
    \bigg\{
    i\xi\ell
    \delta_{\vec{G} +\xi\ell\vec{G}_{n} - \vec{G}'',\: \vec{G}'} 
    - \\ &\mbox{}\hspace{110pt}-
    \sum_{\vec{G}'''}
    (\vec{g}_{n}+\vec{Q})\vec{u}_{\vec{G}'''}
    \delta_{\vec{G} +\xi\ell\vec{G}_{n} - \vec{G}''+\vec{G}''',\: \vec{G}'} 
    \bigg\}.
\end{split}
\end{align}
Additionally, there is a two-phonon vertex that plays a crucial role in the stability of the phason mode, as discussed below (see Fig. \ref{fig:figS3}), but to which the optical conductivity is impervious with our approximations:
\small
\begin{align}\label{eqn:twophonon}
\begin{split}
    \mathcal{g}_{n_{1},n_{2}}^{(\mu,\mu;\xi)}(\vec{k}_{1},\vec{k}_{2}) = 
    -\sum_{\alpha,\alpha',\ell=\pm 1}\sum_{\substack{\vec{G},\vec{G}'\\ \vec{G}'',\vec{G}'''}} \sum_{n=0,1,2}
    &[\mathcal{U}_{n_{2}}^{\vec{G}',\ell,\alpha}(\xi,\vec{k}_{2})]^{*}\:
    [\hat{T}_{\xi}^{(n)}]_{\alpha,\alpha'}
    \mathcal{U}_{n_{1}}^{\vec{G},\Bar{\ell},\alpha'}(\xi,\vec{k}_{1}) 
    \cdot
    \\
    &\left[(\vec{g}_{n}+\vec{Q})
    \delta\vec{\phi}^{(\mu)}_{-\vec{G}''}\left(\vec{k}\right)\right]
    \left[(\vec{g}_{n}+\vec{Q})
    \delta\vec{\phi}^{(\mu)}_{-\vec{G}'''}\left(\vec{k}'\right)\right]
    \delta_{\vec{k}_{2}-\vec{k}_{1}+\vec{G}', \:\vec{k}+\vec{k}' +\vec{G}+\xi\ell\vec{G}_{n}-\vec{G}''-\vec{G}'''}.
\end{split}
\end{align}
\normalsize
Note that for the processes involved in our calculations $\vec{k}_{2}=\vec{k}_{1}$. 

\subsection{Electron-phason coupling from the continuum model}
Let us compare the aforementioned couplings with those originating from the coordinate shift given in Eq.~\eqref{eqn:soliton_coords}. The interlayer tunneling of electrons in the latter case is modified to
\begin{align}\label{eqn:phasonT}
\begin{split}
    \hat{\mathcal{T}}_{\xi}[\vec{\phi}_{0}(\vec{r}-\vec{\zeta})]  &\approx
    \sum_{n=0,1,2}
    e^{i\xi(\vec{G}_{n}+\vec{q}_{0})\vec{r}} \{1 + i\xi(\vec{g}_{n}+\vec{Q})\vec{u}_{0}(\vec{r}) -i\xi(\vec{g}_{n}+\vec{Q})\theta\vec{\hat{z}}\times\vec{\zeta} - i\xi(\vec{g}_{n}+\vec{Q})\zeta_{j}\partial_{j}\vec{u}_{0}(\vec{r})+... \}\hat{T}_{\xi}^{(n)}
    .
\end{split}
\end{align}
Correspondingly (similar for $\partial/\partial(-\zeta_{y})$), 
\begin{align}\label{eqn:electron-zeta}
\begin{split}
    \frac{\partial\hat{\mathcal{T}}_{\xi}}{\partial\zeta_{x}}\bigg|_{\vec{\zeta}=\vec{0}} 
    &=
    \sum_{n=0,1,2} e^{i\xi(\vec{G}_{n}+\vec{q}_{0})\vec{r}} \{-i\xi\theta(\vec{g}_{n}+\vec{Q})_{y} + \xi\sum_{\vec{G}\neq\vec{0}}(\vec{g}_{n}+\vec{Q})\vec{u}_{\vec{G}}G_{x} e^{i\vec{G}\vec{r}}+...\} \hat{T}_{\xi}^{(n)}.
\end{split}
\end{align}
We have checked numerically that, fixing the basis $\{\phi_{\text{ac},x}, \phi_{\text{ac},y}\}$, Eq.~\eqref{eqn:S12} is satisfied, thus the electron-moir\'e phonon coupling in Eq. \eqref{eqn:intercoupling} for the acoustic branches is equivalent to the electron-$\vec{\zeta}$ coupling.
Note, however, that this form of the electron-$\vec{\zeta}$ coupling necessarily requires the relaxation field $\vec{u}_{0}(\vec{r})$ to appear explicitly in the electronic Hamiltonian (last term of Eq. \eqref{eqn:phasonT}), independent of the chosen value of the parameters.
If it did not,  we would not be able to reproduce electron-$\vec{\zeta}$ analytical results using acoustic moir\'e phonons, the reason being that the nonzero addends with $\vec{G}''\neq\vec{0}$ in Eq. \eqref{eqn:intercoupling} would not have a $\vec{\zeta}$-counterpart in Eq.~\eqref{eqn:electron-zeta}. 
This is decisive in order to capture the quantization of the phason response, as shown in Fig. 4 of the main text: We need the electronic wavefunctions of $\hat{H}[\theta\vec{\hat{z}}\times\vec{r} + \vec{u}_{0}(\vec{r})]$, not those with $\vec{u}_{0}(\vec{r})=\vec{0}$ and renormalized model parameters.

\subsection{Damping of moir\'e phonon modes}
We prove that the moir\'e phonon modes of energy smaller than the optical gap -when the flat bands are fully occupied or emptied- are not damped via EPC, as the production of electron-hole pairs is halted by Pauli's principle.

Consider the \textit{paramagnetic} and \textit{diamagnetic} self-energy terms in the Dyson equation, for a branch $\mu$, represented in Fig.~\ref{fig:figS3}(b). In the limit of zero momentum transfer:
\begin{align}
    \underset{\vec{q}\rightarrow 0}{\lim} \Pi_{\text{para}}^{(\mu)}(\vec{q},\omega)
    &= 
    \frac{2}{A}\sum_{\vec{k}\in\text{mBZ}}\sum_{\xi}\sum_{n}\sum_{n'}
    |g^{(\mu;\xi)}_{n,n'}(\vec{k},\vec{k})|^{2}\frac{n_{F}(\varepsilon_{n\xi\vec{k}}) - n_{F}(\varepsilon_{n'\xi\vec{k}})}{\hbar\omega + \varepsilon_{n\xi\vec{k}} - \varepsilon_{n'\xi\vec{k}} + i0^{+}}  
    ,
\end{align}
\begin{align}
    \Pi_{\text{diamag}}^{(\mu)} = \frac{2}{A}\sum_{\vec{k}\in\text{mBZ}}\sum_{\xi}\sum_{n}\mathcal{g}^{(\mu,\mu;\xi)}_{n,n}(\vec{k},\vec{k}) n_{F}(\varepsilon_{n\xi\vec{k}})
    ,
\end{align}
where in the first line we have used $g^{(\mu;\xi)}_{n',n}(\vec{k},\vec{k}) = [g^{(\mu;\xi)}_{n,n'}(\vec{k},\vec{k})]^{*}$.

The diamagnetic term is real. For any energy $\hbar\omega$ within the electronic gap, 
\begin{align}
    \underset{\vec{q}\rightarrow 0}{\lim}\enspace \text{Im}\: \Pi_{\text{para}}^{(\mu)}(\vec{q},\omega) = 
    \frac{2\pi}{A}\sum_{\vec{k}\in\text{mBZ}}\sum_{\xi}\sum_{n}\sum_{n'}
    |g^{(\mu;\xi)}_{n,n'}(\vec{k},\vec{k})|^{2}
    (n_{F}(\varepsilon_{n\xi\vec{k}}) - n_{F}(\varepsilon_{n'\xi\vec{k}}))
    \delta(\hbar\omega + \varepsilon_{n\xi\vec{k}} - \varepsilon_{n'\xi\vec{k}}) = 0.
\end{align}

\subsection{Phason and acoustic modes self-energies}
We now address the question of the stability of phasons in the presence of EPC, which should remain gapless in any continuum model that does not break Goldstone's theorem. 
We prove a similar result as in \cite{Watanabe2014}, but from a different approach, and discuss the fate of acoustic modes.
Let's start by noting that 
\begin{align}
    \hat{\mathcal{H}}_{\xi}[\vec{\phi}_{0}(\vec{r}-\vec{\zeta})] \ket{n\xi\vec{k}} = \varepsilon_{n\xi\vec{k}}\ket{n\xi\vec{k}}
    \Rightarrow
    \frac{\partial}{\partial\zeta_{x}}\ket{n\xi\vec{k}} =
    \sum_{n'\neq n} \frac{\bra{n'\xi\vec{k}} \frac{\partial\hat{\mathcal{H}}_{\xi}[\vec{\phi}_{0}(\vec{r}-\vec{\zeta})]}{\partial\zeta_{x}} \ket{n\xi\vec{k}}}{\varepsilon_{n\xi\vec{k}}-\varepsilon_{n'\xi\vec{k}}}\ket{n'\xi\vec{k}}.
\end{align}
Then, from Hellmann-Feynman theorem and taking an additional derivative with respect to the normal-mode coordinate (drop the $[\vec{\phi}_{0}(\vec{r})]$ from the notation, and take into account that the matrix elements involved fix the phason quasimomentum to $\Gamma$),
\begin{align}\label{eqn:gen_eff_mass}
\begin{split}
    \frac{\partial^{2}\varepsilon_{n\xi\vec{k}}}{\partial\zeta_{x}\partial\zeta_{x}}\bigg|_{\vec{\zeta}=\vec{0}}
    &=
    \bra{n\xi\vec{k}} \frac{\partial\hat{\mathcal{H}}_{\xi}}{\partial\zeta_{x}\partial\zeta_{x}}\ket{n\xi\vec{k}}
    + 
    \sum_{n'\neq n} \frac{\bra{n\xi\vec{k}} \frac{\partial\hat{\mathcal{H}}_{\xi}}{\partial\zeta_{x}} \ket{n'\xi\vec{k}}
    \bra{n'\xi\vec{k}} \frac{\partial\hat{\mathcal{H}}_{\xi}}{\partial\zeta_{x}} \ket{n\xi\vec{k}}  
    +  \bra{n\xi\vec{k}} \frac{\partial\hat{\mathcal{H}}_{\xi}}{\partial\zeta_{x}} \ket{n'\xi\vec{k}}\bra{n'\xi\vec{k}} \frac{\partial\hat{\mathcal{H}}_{\xi}}{\partial\zeta_{x}} \ket{n\xi\vec{k}}}{\varepsilon_{n\xi\vec{k}} - \varepsilon_{n'\xi\vec{k}}}.
\end{split}
\end{align}
This result is akin to the effective-mass theorem in solid state physics.
Using that the single-electron dispersion is invariant under long-wavelength phason distorsions, the left-hand side is exactly zero, whence (applying the same reasoning as below to the $\boldsymbol{\zeta}$ self-energy) it follows that the mode remains gapless.  

On the other hand, to we rewrite Eq.~\eqref{eqn:gen_eff_mass} in terms of acoustic moir\'e phonons, we have to use  Eq.~\eqref{eqn:cond_adh} taking into account missing terms in the two-phonon coupling that come from expanding Eq.~\eqref{eqn:soliton_coords} to higher order in $\vec{\zeta}$, the lowest of which is $\sim|\nabla^{2}\vec{u}_{0}(\vec{r})|\sim \eta\sqrt{1-\eta^{2}}/d$.
In that case, 
\begin{align}
    \mathcal{g}^{(\text{ac},\text{ac};\xi)}_{n,n}(\vec{k},\vec{k}) = 
    -2\sum_{n'\neq n}\frac{g^{(\text{ac};\xi)}_{n,n'}(\vec{k},\vec{k})g^{(\text{ac};\xi)}_{n',n}(\vec{k},\vec{k})}{\varepsilon_{n\xi\vec{k}} - \varepsilon_{n'\xi\vec{k}}} 
    + \mathcal{O}(\eta\sqrt{1-\eta^{2}}\:w_{AB}).
\end{align}
The \textit{diamagnetic} self-energy at  is then given by (see Fig. \ref{fig:figS3} for a diagrammatic representation)
 \begin{align}
     \Pi_{\text{diamag}} = \frac{2}{A}\sum_{\vec{k}\in\text{mBZ}}\sum_{\xi}\sum_{n}\mathcal{g}^{(\text{ac},\text{ac};\xi)}_{n,n}(\vec{k},\vec{k}) n_{F}(\varepsilon_{n\xi\vec{k}})
     \approx
     -\frac{4}{A}\sum_{\vec{k}\in\text{mBZ}}\sum_{\xi}\sum_{n}\sum_{n'\neq n} 
     \frac{g^{(\text{ac};\xi)}_{n,n'}(\vec{k},\vec{k})g^{(\text{ac};\xi)}_{n',n}(\vec{k},\vec{k})}{\varepsilon_{n\xi\vec{k}} - \varepsilon_{n'\xi\vec{k}}} 
     n_{F}(\varepsilon_{n\xi\vec{k}}),
 \end{align}
where the first factor of 2 accounts for spin degeneracy and $n_{F}$ denotes the equilibrium Fermi distribution.
On the other hand, in the limit of zero frequency and momentum transfer, the \textit{paramagnetic} self-energy is
\begin{align}
\begin{split}
    \underset{\substack{\vec{q}\rightarrow 0 \\ \omega\rightarrow 0}}{\lim} \Pi_{\text{para}}(\vec{q},\omega)
    &= 
    \frac{2}{A}\sum_{\vec{k}\in\text{mBZ}}\sum_{\xi}\sum_{n}\sum_{n'}
    g^{(\text{ac};\xi)}_{n,n'}(\vec{k},\vec{k})g^{(\text{ac};\xi)}_{n',n}(\vec{k},\vec{k})\frac{n_{F}(\varepsilon_{n\xi\vec{k}}) - n_{F}(\varepsilon_{n'\xi\vec{k}})}{\varepsilon_{n\xi\vec{k}} - \varepsilon_{n'\xi\vec{k}}} 
    = \\ &=
    \frac{4}{A}\sum_{\vec{k}\in\text{mBZ}}\sum_{\xi}\sum_{n}\sum_{n'\neq n}
    \frac{g^{(\text{ac};\xi)}_{n,n'}(\vec{k},\vec{k})g^{(\text{ac};\xi)}_{n',n}(\vec{k},\vec{k})}{\varepsilon_{n\xi\vec{k}} - \varepsilon_{n'\xi\vec{k}}} n_{F}(\varepsilon_{n\xi\vec{k}}).
\end{split}
\end{align}
In the last line we have relabeled the dummy indices $n,n'$ of one of the Fermi distributions, and we have taken into account that the term $n=n'$ vanishes, so it can be safely removed from the sum. Thus,
\begin{align}
    \Pi_{\text{diamag}} + \underset{\substack{\vec{q}\rightarrow 0 \\ \omega\rightarrow 0}}{\lim} \Pi_{\text{para}}(\vec{q},\omega) \approx 0.
\end{align}
These are all the possible single-loop diagrams contributing to the moir\'e phonon self-energy.

In closing, let us stress that this result becomes exact for the phason variable $\vec{\zeta}$, whereas the self-energy calculation in the stacking variables receives contributions of order $\mathcal{O}(\eta\sqrt{1-\eta^{2}}\:w_{AB})$ in the $\mathbf{q}\rightarrow 0$ limit.

\section{Conductivity calculation}
\subsection{Optical f-sum rule}
The optical f-sum rule holds in the presence of interactions. Hence, it must be true that the phonon contribution to the optical conductivity satisfies
\begin{align}\label{eqn:fsumrule}
    \int_{0}^{\infty}d\omega\: \mathrm{Re}\: \sigma_{xx}^{(1)}(\omega) = 0.
\end{align}
We now explicitly prove that the diagrams that we compute exhaust this condition. Assume in what follows the limit $\vec{k}\rightarrow\vec{0}$ and drop the momentum arguments. 
Given that $\chi_{j_{x}\phi_{\alpha,y}}(\omega)$, $\chi_{\phi_{\alpha,y} j_{x}}(\omega)$, and $D_{\alpha}(\omega)$ are causal response functions, they satisfy Kramers-Kronig relations:
\begin{align}
    \text{Re} \chi_{j_{x}\phi_{\alpha,y}}(\omega) = \frac{1}{\pi}\mathcal{P}\int_{-\infty}^{\infty} \frac{\text{Im}\:\chi_{j_{x}\phi_{\alpha,y}}(\omega')}{\omega'-\omega} d\omega',
    &&
    \text{Im} \chi_{j_{x}\phi_{\alpha,y}}(\omega) = -\frac{1}{\pi}\mathcal{P}\int_{-\infty}^{\infty} \frac{\text{Re}\:\chi_{j_{x}\phi_{\alpha,y}}(\omega')}{\omega'-\omega} d\omega'.
\end{align}
Analogously for $\chi_{\phi_{\alpha,y},j_{x}}(\omega)$ and $D_{\alpha}$. To simplify the notation, let's define the complex function $\tilde{\chi}_{\alpha}:= \frac{1}{\omega}\chi_{j_{x}\phi_{\alpha,y}}\chi_{\phi_{\alpha,y} j_{x}}$, which is also analytic in the upper-half plane. Now,
\begin{align}
    \sigma^{(1)}_{xx}(\omega) = \frac{i}{\omega}\sum_{\alpha\in E_{1}}\chi_{j_{x}\phi_{\alpha,y}}(\omega)D_{\alpha}(\omega)\chi_{\phi_{\alpha,y} j_{x}}(\omega) 
    = i \sum_{\alpha\in E_{1}}\tilde{\chi}_{\alpha}(\omega)D_{\alpha}(\omega).
\end{align}
The real part of a causal response function is an even function of the frequency, whence
\begin{align}
\begin{split}
    \int_{0}^{\infty}d\omega\: \text{Re}\:\sigma^{(1)}_{xx}(\omega) &=
    \frac{1}{2}\int_{-\infty}^{\infty}d\omega\: \text{Re}\:\sigma^{(1)}_{xx}(\omega) =
    -\frac{1}{2}\sum_{\alpha\in E_{1}}\int_{-\infty}^{\infty}d\omega\: \text{Re}\:\tilde{\chi}_{\alpha}(\omega)\text{Im}\:D_{\alpha}(\omega) + \text{Im}\:\tilde{\chi}_{\alpha}(\omega)\text{Re}\:D_{\alpha}(\omega)
    =\\ \\ &=
    \frac{1}{2\pi}\sum_{\alpha\in E_{1}}\int_{-\infty}^{\infty}d\omega\int_{-\infty}^{\infty}d\omega' \frac{ \text{Re}\:\tilde{\chi}_{\alpha}(\omega)\text{Re}\:D_{\alpha}(\omega')}{\omega'-\omega }
    +  \frac{ \text{Re}\:\tilde{\chi}_{\alpha}(\omega')\text{Re}\:D_{\alpha}(\omega)}{\omega'-\omega }
    =\\ \\ &=
    \frac{1}{2\pi}\sum_{\alpha\in E_{1}}\int_{-\infty}^{\infty}d\omega\int_{-\infty}^{\infty}d\omega' \frac{ \text{Re}\:\tilde{\chi}_{\alpha}(\omega)\text{Re}\:D_{\alpha}(\omega')}{\omega'-\omega }
    +  \frac{ \text{Re}\:\tilde{\chi}_{\alpha}(\omega)\text{Re}\:D_{\alpha}(\omega')}{\omega-\omega'}
    = 0.
\end{split}
\end{align}

\subsection{Sliding  conductivity}
This section is devoted to the derivation of Eqs. (14) and (15) in the main text. We are interested in the acoustic branches, hence we drop the mode index. For the second diagram of the optical conductivity, we have
\begin{align}\label{eqn:sigmaphxx}
\begin{split}
    \sigma_{xx}^{\text{ph}}(\omega) &=
    \underset{\vec{k}\rightarrow 0 }{\lim}\: \frac{i}{\omega} 
    \chi_{j_{x}(\vec{k})\phi_{y}(-\vec{k})}(\omega) D_{\text{ph}}(\vec{k},\omega) \chi_{\phi_{y}(\vec{k})j_{x}(-\vec{k})}(\omega)
    = \\ &=
    \frac{i}{\omega} \left|\frac{2e}{A}\sum_{n_{1},n_{2}}\sum_{\xi}\sum_{\vec{k}\in\text{mBZ}} v_{x;n_{2}n_{1}}^{(\xi)}(\vec{k},\vec{k}) \:
    g_{n_{1}n_{2}}^{(\text{ph};\xi)}(\vec{k},\vec{k}) 
    \frac{n_{F}(\varepsilon_{n_{2}\xi\vec{k}}) - n_{F}(\varepsilon_{n_{1}\xi\vec{k}})}{\hbar\omega + \varepsilon_{n_{2}\xi\vec{k}} - \varepsilon_{n_{1}\xi\vec{k}} +i0^{+}} \right|^{2}
    \frac{2}{\varrho}\frac{1}{\omega^{2} - \omega^{2}_{\vec{0}} + i\gamma\omega},
\end{split}
\end{align}
where the matrix elements of the velocity operator (diagramatically represented in Fig.~3(a) of the main text) are given by
\begin{align}
    v_{x;n_{1}n_{2}}^{(\xi)}(\vec{k}_{1},\vec{k}_{2})
    =\frac{1}{\hbar}
    \bra{n_{2},\xi,\vec{k}_{2}}\frac{\partial \hat{\mathcal{H}}_{\xi}}{\partial k_{x}}\ket{n_{1},\xi,\vec{k}_{1}}.
\end{align}
The interlayer and relaxation effects are contained in the eigenstates $\ket{n,\xi,\vec{k}}$, since $\partial \hat{\mathcal{H}}_{\xi} / \partial k_{x}$ is diagonal in layer and takes the same expression (within each layer) as in graphene. 

To proceed further, note that the contribution to Eq.~\eqref{eqn:sigmaphxx} of the first term in the decomposition
\begin{align}
    \frac{1}{\hbar\omega + \varepsilon_{n_{2}\xi\vec{k}} - \varepsilon_{n_{1}\xi\vec{k}} +i0^{+}} = 
    \frac{1}{\varepsilon_{n_{2}\xi\vec{k}} - \varepsilon_{n_{1}\xi\vec{k}}} 
    - \frac{\hbar\omega}{(\varepsilon_{n_{2}\xi\vec{k}} - \varepsilon_{n_{1}\xi\vec{k}})^{2}}
    + \frac{(\hbar\omega)^{2}}{(\varepsilon_{n_{2}\xi\vec{k}} - \varepsilon_{n_{1}\xi\vec{k}})^{2} (\hbar\omega + \varepsilon_{n_{2}\xi\vec{k}} - \varepsilon_{n_{1}\xi\vec{k}} + i0^{+})}
\end{align}
vanishes, for the current and phonon interaction vertices transform differently under time-reversal symmetry. 
For the same reason, the low-frequency mixed polarization bubble coming from the second term in this decomposition is purely imaginary, hence the optical conductivity of the phason at zero temperature can be approximated as
\begin{subequations}
\begin{align}
    \sigma_{xx}^{\text{ph}}(\omega) &\approx
    \frac{2\omega}{\varrho}
    \left|\frac{2e\hbar}{A}\sum_{n_{1},n_{2}}\sum_{\xi}\sum_{\vec{k}\in\text{mBZ}}  \frac{v_{x;n_{2}n_{1}}^{(\xi)}(\vec{k},\vec{k}) \:g_{n_{1}n_{2}}^{(\text{ph};\xi)}(\vec{k},\vec{k})}{(\varepsilon_{n_{2}\xi\vec{k}} - \varepsilon_{n_{1}\xi\vec{k}})^{2}} 
    (n_{F}(\varepsilon_{n_{2}\xi\vec{k}}) - n_{F}(\varepsilon_{n_{1}\xi\vec{k}}))
    \right|^{2}
    \frac{1}{\gamma\omega -i(\omega^{2} - \omega^{2}_{\vec{0}})}
    = \\ &=
    \frac{2\eta^{2}}{\theta^2\varrho}
    \left|\frac{e}{A}\sum_{n_{1}}^{\text{occ}}\sum_{\xi}\sum_{\vec{k}\in\text{mBZ}}\Omega^{(n)}_{q_{x},\zeta_{x}}(\vec{k}) 
    \right|^{2}
    \frac{1}{\gamma -i(\omega - \frac{\omega^{2}_{\vec{0}}}{\omega})}
    = 
    \frac{2\eta^{2}n^{2}e^{2}\gamma^{-1}}{\theta^{2}\varrho}
    \frac{1}{1 -i\gamma^{-1}(\omega - \frac{\omega^{2}_{\vec{0}}}{\omega})},
\end{align}
\end{subequations}
whence we can readily obtain the effective mass $m^{*} = \frac{\theta^{2}\varrho}{2\eta^{2}n}$.

\subsection{Numerical implementation}
In the diagonalization of the electronic Hamiltonian we use 10 stars and a sample of 2700 points in the moir\'e Brillouin zone.
A star consists of 6 moir\'e reciprocal lattice vectors related by $C_{6z}$ rotations. 
We independently minimize the mechanical free energy iteratively and diagonalize the corresponding dynamical matrix, as in \cite{Ochoa2025}, using 55 stars and sampling the real-space moir\'e unit cell with 6400 points.

In the EPC calculations, the phonon eigenvectors are cut-off (and then normalized to 1) to match the length of the electronic wavefunctions. For the calculations of the optical conductivity shown in the main text, we keep 10 bands below the Fermi level and 10 bands above. We have checked the convergence of the results with respect to the number of bands including up to 60 (30 occupied, 30 unoccupied). 

For each twist angle, we calculate $\sigma^{(0)}_{xx}(\omega)$ over 250 equally distributed frequencies with a broadening of 0.5 meV (1 meV) at $\theta=1.05^{\circ}$ ($\theta=1.54^{\circ}$, $2^{\circ}$). A broadening of 3 meV is needed at $\theta=2^{\circ}$ and charge neutrality to avoid numerical oscillations in the optical absorption. 
On the other hand, for each of the four moir\'e phonon peaks shown in the figures, $\sigma^{(1)}_{xx}(\omega)$ is computed over 100 frequencies around the energy of the mode.
The damping constant of the moir\'e phonons is determined  extrapolating a fit to transport measurements of minimally twisted tBG ($\gamma = 0.465$ meV at $\theta = 0.4^{\circ}$ \cite{Pezzini2025}) as $\gamma\propto\theta^{-3}$  \cite{Fernandes2022}.

\end{document}